\documentclass[11pt]{article}

\usepackage[english]{babel}
\usepackage[utf8]{inputenc}

\usepackage{amssymb,amsmath,amsfonts,mathrsfs}
\usepackage{xcolor}
\usepackage[
    colorlinks=true,
    linkcolor=blue,
    citecolor=blue,
    urlcolor=blue
]{hyperref}

\advance \textheight by \topskip
\usepackage{array}
\usepackage{authblk}
\usepackage{relsize}

\def\be{\begin{equation}}
\def\ee{\end{equation}}
\def\bea{\begin{eqnarray}}
\def\eea{\end{eqnarray}}
\def\beq{\begin{equation}}
\def\eeq{\end{equation}}
\def\beqa{\begin{eqnarray}}
\def\eeqa{\end{eqnarray}}
\def\Epsilon{\varepsilon}

\newcommand{\I}{\mathbf{1}}

\newcommand{\so}{\mathfrak{so}}

\newcommand{\N}{\mathbf{N}}

\newcommand{\CJ}{\mathcal{J}}
\newcommand{\CK}{\mathcal{K}}

\newcommand{\CC}{\mathcal{C}}

\newcommand{\Dt}{{D_t^{(\Epsilon)}}}
\newcommand{\ie}{{\textit{i.e.}}}
\newcommand{\cf}{{\textit{cf.}}}

\begin{document}

\title{\textbf{Carrollian Wave Equations for Arbitrary Spin:
Anyons and the Exotic Particle on the Noncommutative Plane
  }}

\author[1,2]{Mauricio Valenzuela%
\thanks{\href{mailto:mauricio.valenzuela@uss.cl}{mauricio.valenzuela@uss.cl}}}

\affil[1]{Centro de Estudios Cient\'{\i}ficos (CECS), Avenida Arturo Prat 514, Valdivia, Chile}
\affil[2]{Facultad de Ingenier\'{\i}a, Universidad San Sebasti\'an, sede Valdivia, General Lagos 1163, Valdivia
5110693, Chile}

\date{}

\maketitle

\begin{abstract}

We construct the Carrollian limit of the Cort\'es--Plyushchay
wave equations for planar anyons, grading the spin-tower components by
powers of $c$ and taking the limit at fixed rest energy.  The result is a linear system describing Carrollian
particles of any real spin --- \emph{Carrollian anyons} --- and, at
(half-)integer spin, Carrollian bosons, fermions and higher-spin fields.
 The system is a first-order square root of the standard Carroll
wave equation $(\partial_t^2+E^2)\Phi=0$, whose spectrum consists of two flat branches at plus and
minus the rest energy.  Scaling the spin along with $c$, in the Carrollian
analogue of the Jackiw--Nair limit, yields instead the exotic Carroll
algebra, with a second central charge, non-commuting boosts, and
observable coordinates spanning a noncommutative plane.  We
show how the Galilean and Carrollian theories branch from one common
similarity transformation: the Galilean limit expels the
negative-energy modes, whereas the Carroll limit retains both
relativistic branches.
\end{abstract}


\newpage

{
\hypersetup{linkcolor=black}
\tableofcontents
}

%%%%%%%%%%%%%%%%%%%%%%%%%%%%%%%%%%%%%%%%%%%%%%%%%%%%%%%%%%%%%%%%%%%%%%%%%%%%%%
\section{Introduction}
\label{sec:intro}
%%%%%%%%%%%%%%%%%%%%%%%%%%%%%%%%%%%%%%%%%%%%%%%%%%%%%%%%%%%%%%%%%%%%%%%%%%%%%%

Sending the speed of light to infinity degrades relativistic physics into
the familiar Galilean world; sending it to zero produces instead the
Carrollian world \cite{Levy-Leblond:1965dsc,SenGupta:1966qer,Bacry:1968zf},
in which light cones collapse onto the time axis and dynamics becomes
ultra-local in space.  Long dismissed as a kinematical curiosity, Carroll
symmetry has turned out to be the intrinsic symmetry of null hypersurfaces
\cite{Duval:2014uoa,Ciambelli:2019lap} --- black-hole horizons among them
\cite{Donnay:2019jiz,Ecker:2023uwm} --- and to underlie flat-space
holography \cite{Donnay:2022aba,Mason:2023mti,Alday:2024yyj}, Carrollian
hydrodynamics \cite{deBoer:2021jej,Marsot:2022imf} and tensionless strings
\cite{Isberg:1993av,Bagchi:2015nca}; the ultralocal field theories studied
long ago by Klauder \cite{Klauder:1970cs} were Carrollian \emph{avant la
lettre}.  At the level of particles and wave equations, Carrollian dynamics
has been explored for planar Carroll particles and the associated quantum
equation \cite{Marsot:2021tvq}, for Carroll/fracton particles
\cite{Figueroa-OFarrill:2023qty}, and at one loop for Carroll scalars
\cite{Banerjee:2023jpi}, while systematic routes to Carrollian and Galilean
field theories include $c$-power field redefinitions, null reductions,
field expansions, seed Lagrangians and intrinsically geometric
constructions
\cite{Bergshoeff:2015sic,Bagchi:2016bcd,Bergshoeff:2017btm,Hansen:2021fxi,
deBoer:2021jej,Bergshoeff:2022eog}.  Carroll fermions, in particular, have
attracted considerable recent attention
\cite{Bagchi:2019xfx,Bergshoeff:2023vfd,Koutrolikos:2023evq,Ekiz:2025hdn,Parekh:2026wri},
while Carrollian higher-spin algebras and field theories have been
constructed in conformal and null-infinity settings
\cite{Campoleoni:2021blr,Liu:2023jnc}.

An independent line of development concerns \emph{anyons}: in $2+1$
dimensions spin is not restricted to (half-)integer values, and identical
particles need not be bosons or fermions.  This follows from the topology of
their planar configuration space
\cite{Leinaas:1977fm,Wilczek:1982wy,Wu:1984hj} and, group-theoretically, from
the existence of arbitrary real-spin representations of the planar Lorentz
and Poincar\'e groups.  Far from being
an academic curiosity, fractional statistics is realised by the
quasi-particle excitations of the fractional quantum Hall effect
\cite{Tsui:1982yy,Laughlin:1983fy,Halperin:1984fn,Arovas:1984qr}, by now
confirmed by direct experimental observations of fractional charge and of
anyonic braiding phases
\cite{dePicciotto:1997qc,Bartolomei:2020qfr,Nakamura:2020mok}.  Interest
in anyons has been further amplified by their applied, quantum-information
side: non-Abelian anyons provide the hardware model of topological quantum
computation \cite{Kitaev:1997wr,Kitaev:2005hzj,Nayak:2008zza}, and
non-Abelian braiding has recently been implemented on superconducting
quantum processors \cite{Andersen:2022xmz}.  On the theoretical side,
fractional-spin fields have been coupled to higher-spin gravity in the
Chern--Simons formulation \cite{Boulanger:2013naa,Boulanger:2015uha},
suggesting that higher-spin (topological) interactions may be relevant to quantum Hall
phenomenology.

A covariant, and remarkably economical, field-theoretical description of
free relativistic anyons is provided by the vector system of wave equations
of Cort\'es and Plyushchay \cite{Cortes:1992fa},
\be
\label{intro:V}
V_\mu\psi=0\,,\qquad
V_\mu=\alpha\, P_\mu-i\epsilon_{\mu\nu\lambda}P^{\nu}\CJ^{\lambda}
+mc\,\CJ_\mu\,,
\ee
where $\CJ_\mu$ generates a half-bounded (in general infinite-dimensional)
representation of the planar Lorentz algebra $\so(2,1)$ of spin $\alpha$.
For $\alpha=-j$ with $j$ a (half-)integer, \eqref{intro:V} describes
finite $(2j+1)$-component Poincar\'e representations, including
Dirac fields ($j=1/2$) and
topologically massive electrodynamics ($j=1$) \cite{Deser:1981wh}; for
non-(half-)integer $\alpha$ it describes fractional-spin particles.  It
belongs to the broader family of infinite-component anyon equations that
includes the Jackiw--Nair and Majorana--Dirac formulations
\cite{Jackiw:1990ka,Plyushchay:1990cv,Plyushchay:1991qd,
Plyushchay:1990rt}, as well as the quartion and semion constructions
\cite{Volkov:1989qa,Sorokin:1992sy}.

There are, in fact, two essentially different infinite-dimensional anyon
sectors \cite{Horvathy:2010vm}.  The conventional one is based on unitary
discrete-series representations of the Lorentz algebra, whereas the second
uses half-bounded non-unitary representations.  Only the latter sector
interpolates continuously between the finite-dimensional representations
of ordinary bosons and fermions.  The collapse of the infinite spin tower
to a finite one is beautifully encoded by Verma modules: at
$\alpha=-j$, all states beyond the first $2j+1$ form a null invariant
submodule, while $|2j)$ becomes the highest-weight state of the finite
quotient.  The three possibilities are summarised in
Table~\ref{tab:LorentzClasses}.

\begin{table}[t]
\centering
\renewcommand{\arraystretch}{1.2}
\begin{tabular}{@{}>{\centering\arraybackslash}p{0.19\textwidth}
>{\raggedright\arraybackslash}p{0.42\textwidth}
>{\raggedright\arraybackslash}p{0.25\textwidth}@{}}
\hline
Parameter & Lorentz representation & Particle sector \\
\hline
$\alpha>0$ & infinite-dimensional unitary discrete series $D^+_\alpha$
& unitary anyons \\
$\alpha=-j$ & finite-dimensional non-unitary $\widetilde D^j$
& ordinary bosons and fermions \\
$-j<\alpha<-j+\tfrac12$ & infinite-dimensional half-bounded non-unitary
$\widetilde D^+_\alpha$ & boson--fermion-interpolating anyons \\
\hline
\end{tabular}
\caption{Lowest-weight Lorentz representations admitted by
\eqref{intro:V}, with $j=\tfrac12,1,\tfrac32,\ldots$.  The corresponding
highest-weight representations describe the opposite spin orientation.}
\label{tab:LorentzClasses}
\end{table}

The Galilean limit of these equations is well understood
\cite{Horvathy:2004fw,Horvathy:2006pw,Horvathy:2010vm}.  A similarity
transformation grading the components of the spin tower by powers of $c$,
followed by $c\to\infty$ at fixed mass, yields a universal first-order
system whose integrability condition is the free Schr\"odinger equation ---
the arbitrary-spin generalisation of the L\'evy-Leblond equations
\cite{Levy-Leblond:1967eic} --- carrying an irreducible representation of the
Bargmann (centrally extended Galilei) algebra with spin $\alpha$.  An
``exotic'' variant of the limit,
in which the spin is scaled together with $c$, produces instead the exotic
Galilei algebra with two central charges and non-commuting boosts
\cite{Lukierski:1996br,Jackiw:2000tz,Duval:2000xr,Horvathy:2004fw,
Horvathy:2006pw}.

The purpose of this paper is to show that the same grading strategy also
admits a non-degenerate Carrollian contraction.  A similarity
transformation grading the entire tower by powers of $c$, followed by
$c\to0$ at fixed rest energy $E=mc^{2}$ (and hence $m\to\infty$), yields a
consistent Carrollian theory for arbitrary real spin.  Since the Lorentz
spin module survives the contraction, each of the three relativistic
sectors in Table~\ref{tab:LorentzClasses} has a direct Carrollian
counterpart.  The resulting first-order equations describe Carrollian
anyons for generic $\alpha$, and finite-component Carrollian bosons,
fermions and higher-spin fields for $\alpha=-j$. Their consistency condition is
\be
\mathcal D_C \Phi=0\,,\qquad \mathcal D_C:=(i\partial_t-E)(i\partial_t+E)\,.
\label{CarrollCasEq}
\ee
This is the massive Carroll scalar equation appearing in
Carroll particle quantisation and field theory
\cite{Bergshoeff:2014jla,Marsot:2021tvq,Ciambelli:2023xqk,
Parekh:2026wri}. Here it follows from the vector system in
 the same sense that the L\'evy--Leblond equations yield the free
Schr\"odinger equation.

A second, genuinely exotic contraction is obtained by scaling the spin as
$\alpha\to\infty$ while keeping $\kappa_C=\alpha c^2$ fixed.  The internal
Lorentz $\mathfrak{sp}(2,\mathbb{R})\cong\so(2,1)$ module then
contracts to an oscillator (Heisenberg algebra) Fock module, and the scaled
spin becomes the exotic central charge in
$[B_i,B_j]=-i\kappa_C\epsilon_{ij}$.  The contracted wave equations thereby
provide a dynamical representation of the exotic Carroll algebra studied
from group-theoretical and coadjoint-orbit viewpoints in
\cite{Marsot:2021tvq}.  The spin-dependent dressing of the
spatial coordinates leads to the boost-defined observables $\mathcal X_i$,
which preserve the physical subspace and span a
noncommutative plane with area scale
$\theta_C=\kappa_C/E^2=i [\mathcal X_1,\mathcal X_2]$, while the exotic wave
constraint ties each momentum
mode to a Glauber coherent state of the internal oscillator.  This gives a
lowest-Landau-level-like Bargmann--Fock realisation of the surviving spin
tower.  The relativistic anyon possesses an observable covariant Foldy--Wouthuysen-type replacement of the canonical coordinates $x^\mu$
that preserves the physical subspace
\cite{Foldy:1949wa,Cortes:1995wa}.  Its temporal component $X^0$ is thus a corrected relativistic time coordinate.  In the exotic Carroll limit the rescaled coordinate
$\mathcal T:=\lim_{c\to0}X^0/c$ remains a finite Hermitian operator rather
than reducing to the parameter $t$ alone.  This contrasts with the Galilean
limit, in which the analogous spin correction vanishes and absolute time
remains undeformed.  Its momentum--spin correction makes it
noncommutative with the spatial observables $\mathcal X_i$, whereas its constraint-adapted complex improvement
$\mathcal T'$ contains, on shell, the imaginary integer-spaced shift
$i\N/E$.  Although $\mathcal T'$ is not a physical time observable of the
fixed-energy system, this formal complex-time ladder is a distinctive remnant
of the exotic contraction.

Thus, our two main results are the arbitrary-spin Carrollian counterpart
of the L\'evy--Leblond square-root construction and a wave-equation
realisation of the exotic Carroll particle on the noncommutative plane.

The paper is organised as follows.  Section~\ref{sec:setup} recalls the Plyushchay-Cortes
relativistic vector system.  Section~\ref{sec:carroll} introduces the
Carroll similarity transformation and derives the Carrollian anyon
equations and their solutions.  Section~\ref{sec:carrollalgebra} derives
their Carroll symmetry, establishes covariance of the wave system, computes
the Casimir operators and classifies the boost sectors according to the
grading exponent.  Section~\ref{sec:examples} works out the finite-dimensional
(half-)integer-spin truncations explicitly: $j=\tfrac12$ yields the
Carrollian L\'evy--Leblond system, $j=1$ the Carroll limit of the
Deser--Jackiw--Templeton topologically massive gauge system, and
$j=\tfrac32$ a finite-energy Carrollian Rarita--Schwinger system; closed
equations and solutions are then given for arbitrary (half-)integer $j$.
Section~\ref{sec:exotic} develops the exotic Carroll wave equation
and algebra, the resulting noncommutative plane and Carrollian time
coordinate, and the two gapless boundaries of the exotic family.
Section~\ref{sec:branching} exhibits how the
Galilean and Carrollian theories branch from one common phase-dressed
similarity transformation, and draws the corresponding lesson on the
fate of the two relativistic energy branches.  We conclude in
Section~\ref{sec:conclusions}.  Appendix~\ref{app:conv} collects
conventions and algebraic identities; Appendix~\ref{app:bigXlimit}
analyses the direct $c\to0$ limit of the relativistic covariant position
operator.

%%%%%%%%%%%%%%%%%%%%%%%%%%%%%%%%%%%%%%%%%%%%%%%%%%%%%%%%%%%%%%%%%%%%%%%%%%%%%%
\section{The relativistic anyon system}
\label{sec:setup}
%%%%%%%%%%%%%%%%%%%%%%%%%%%%%%%%%%%%%%%%%%%%%%%%%%%%%%%%%%%%%%%%%%%%%%%%%%%%%%

Following the notation and conventions of Ref.~\cite{Horvathy:2010vm},
we work with the mostly-plus metric
$\eta_{\mu\nu}=\mathrm{diag}(-1,1,1)$, take
$\epsilon^{012}=1$, and keep the speed of light explicit throughout,
$x^{0}=ct$.  The Poincar\'e generators are
\be
P_t:=i\partial_t=cP^0=-cP_0\,,\qquad
P_\mu=-i\partial_\mu\,,\qquad
\mathcal M_\mu=-\epsilon_{\mu\nu\lambda}x^\nu P^\lambda+\CJ_\mu\,.
\label{Pgen}
\ee
where $\CJ_\mu$ is the spin part to be specified below.  With these
conventions they obey the $(2+1)$-dimensional Poincar\'e algebra
\be
\label{PoincareAlg}
[P_\mu,P_\nu]=0\,,\qquad
[\mathcal M_\mu,P_\nu]=-i\epsilon_{\mu\nu\lambda}P^\lambda\,,
\qquad
[\mathcal M_\mu,\mathcal M_\nu]
=-i\epsilon_{\mu\nu\lambda}\mathcal M^\lambda\,.
\ee
We set $P_\pm=P_1\pm iP_2$,
$x_\pm=x_1\pm ix_2$, and similarly for any other vector, so that
\be
\label{xPcomm}
[x_\pm,P_\mp]=2i\,,\qquad [x_\pm,P_\pm]=0\,.
\ee

The Cort\'es--Plyushchay system \cite{Cortes:1992fa} is \eqref{intro:V} with
$\CJ_\mu$ in a lowest-weight, half-bounded representation of $\so(2,1)$
\cite{Bargmann:1946me,Barut:1965}, whose matrix elements are
\be
\label{Dplus}
\CJ_0|n)=(\mathsmaller{\alpha+n})|n)\,,\qquad
\CJ_+|n)=C^{\alpha}_{n}|\mathsmaller{n+1})\,,\qquad
\CJ_-|n)=C^{\alpha}_{n-1}|n-1)\,,
\ee
with $C^{\alpha}_{n}=\sqrt{(2\alpha+n)(n+1)}$, $n$ a
non-negative integer, and $|n)\propto\CJ_+^n|0)$ the level-$n$
descendants of the lowest-weight state $|0)$, satisfying
$\CJ_-|0)=0$ and $\CJ_0|0)=\alpha|0)$.  For $\alpha>0$, these formulas define the irreducible unitary
discrete series ${\cal D}^{+}_{\alpha}$; for generic $\alpha<0$, they define
the infinite-dimensional non-unitary modules $\widetilde D^+_\alpha$ of
Table~\ref{tab:LorentzClasses}.  At the exceptional values $\alpha=-j$,
with $j$ a positive (half-)integer, $C^{-j}_{2j}=0$ and the representation
truncates to $n=0,\ldots,2j$.  In terms of the underlying lowest-weight
Verma module, a singular vector appears at level $2j+1$ and its descendants
generate a null invariant submodule; quotienting by it leaves the
finite-dimensional module $\widetilde D^j$, with $|2j)$ as its
highest-weight state.  See Appendix~\ref{app:conv} for more details and the corresponding
finite-dimensional conventions.

Expanding $\psi(x)=\sum_n\psi_n(x)|n)$, and linearly combining two equations from \eqref{intro:V}, we obtain
\be
\label{Vpm}
V_\pm\psi=0\,,\qquad  V_\pm:=V_1\pm iV_2=(\alpha \mp \CJ_0)P_\pm +(mc\mp P^{0})\CJ_\pm\,,
\ee
equivalent in components to
\bea
&\sqrt{n+2\alpha}\,(mc-P^{0})\psi_n-\sqrt{n+1}\,P_+\psi_{n+1}=0\,,&
\label{HP1}\\[4pt]
&\sqrt{n+2\alpha}\,P_-\psi_n+\sqrt{n+1}\,(mc+P^{0})\psi_{n+1}=0\,.&
\label{HP2}
\eea
They generate
\be
\label{V0}
V_0\psi=0\,,\qquad V_0=\alpha P_0+mc\,\CJ_0+\tfrac12\bigl(P_-\CJ_+-P_+\CJ_-\bigr)\,,
\ee
as an integrability condition \cite{Cortes:1992fa}.
The system implies the Klein--Gordon equation and the $2+1$-dimensional Pauli--Luba\'nski
condition
\be \label{KGPL}
(P^{2}+m^{2}c^{2})\psi=0\,,\qquad (P\CJ-\alpha mc)\psi=0.
\ee
Hence they determine the Poincar\'e group Casimir operators of mass and spin, respectively
\be \label{PCas}
\mathcal C_1:=P^2\,,\qquad \mathcal C_2:=P\CJ.
\ee
For further details see references \cite{Cortes:1992fa,Horvathy:2006pw,Horvathy:2010vm}.

%%%%%%%%%%%%%%%%%%%%%%%%%%%%%%%%%%%%%%%%%%%%%%%%%%%%%%%%%%%%%%%%%%%%%%%%%%%%%%
\section{The Carrollian anyon wave equation}
\label{sec:carroll}
%%%%%%%%%%%%%%%%%%%%%%%%%%%%%%%%%%%%%%%%%%%%%%%%%%%%%%%%%%%%%%%%%%%%%%%%%%%%%%

\subsection{The Carroll similarity transformation}
\label{sec:SC}

Before taking the limit of the vector system of equations \eqref{intro:V}, it is instructive to see the limit first on the free scalar
field, for which the Klein--Gordon equation $(P^{2}+m^{2}c^{2})\psi=0$
reads, after multiplication by $c^{2}$,
\be
\label{KGscalar}
\Bigl(\partial_t^{2}-c^{2}\vec\nabla^{2}+m^{2}c^{4}\Bigr)\psi=0\,.
\ee
Taking $c\to0$ at fixed mass requires no field redefinition, but it is
degenerate: both the gradient and the rest-energy terms drop, leaving
$\partial_t^{2}\psi=0$ --- a frozen particle of exactly zero energy.
Hence, the naive limit yields a static vanishing energy system, which has little interest to us.
Taking instead $c\to0$ at \emph{fixed rest energy},
\be
\label{Clim}
c\to0\,,\qquad m\to\infty\,,\qquad E:=mc^{2}=\text{fixed}\,,
\ee
the rest-energy term survives, $m^{2}c^{4}=E^{2}$, and \eqref{KGscalar}
tends to the Carroll equation $(\partial_t^{2}+E^{2})\psi=0$,
a momentum-independent dispersion with the two energy branches $\pm E$.

The operator $\partial_t^{2}$ is a Casimir invariant of the planar
Carroll group, and the Carroll equation fixes its eigenvalue to $-E^2$
\cite{Marsot:2021tvq}.  The same equation follows from quantising the
massive Carroll particle \cite{Bergshoeff:2014jla} and appears as the
free massive scalar equation in Carrollian field theory
\cite{Mehra:2023rmm,Banerjee:2023jpi,Sharma:2025rug}; see also the
intrinsic treatment of Carrollian scalar dynamics in
\cite{Ciambelli:2023xqk} and the fixed-rest-energy two-branch structure
of \cite{Parekh:2026wri}.  Historically, the defining absence of spatial
gradients --- independent oscillator dynamics at each spatial point ---
already appeared in Klauder's ultralocal scalar models
\cite{Klauder:1970cs}. In the Galilean limit, by contrast, a time-dependent phase redefinition
is required to control the divergent rest-energy term.  The comparison
of the two regimes is deferred to
Section~\ref{sec:branching}.

For a field with non-trivial spin the naive limit is not sufficient:
the components of the spin tower must be rescaled
\emph{non-homogeneously}, in a manner consistent with the limit, so
that the spin content survives the contraction --- otherwise the spin
information is lost, \cf\ the degenerate central class of
Section~\ref{sec:classes}.  We therefore introduce the similarity
transformation adapted to the Carroll limit,
\be
\label{SC}
\Phi:=S\,\psi\,,\qquad
S:=c^{\,\N}\,,\qquad
\N:=\CJ_0-\alpha\,,\qquad \N|n)=n|n)\,,
\ee
that is, $\phi_n=c^{\,n}\,\psi_n$; in matrix form
$c^{\,\N}=\mathrm{diag}(1,c,c^{2},\dots)$.
The transformation \eqref{SC} induces on any operator $\mathfrak O$ the
similarity map $\widetilde{\mathfrak O}=S\,\mathfrak O\,S^{-1}$.
Since $S$ depends only on $\N=\CJ_0-\alpha$, the orbital operators are
untouched, $\widetilde x_i=x_i$, $\widetilde P_i=P_i$,
$\widetilde P^{0}=P^{0}=P_t/c$, while on the spin
generators it acts as
\be
\label{tilCJ}
\widetilde\CJ_0=\CJ_0\,,\qquad
\widetilde\CJ_+=c\,\CJ_+\,,\qquad
\widetilde\CJ_-=\frac1c\,\CJ_-\,.
\ee
Applying \eqref{tilCJ} to \eqref{Vpm} and \eqref{V0}, and using
$c\,(mc-P^{0})=-(P_t-E)$ and $c\,(mc+P^{0})=P_t+E$, we
obtain the \emph{exact} transformed vector operators
\bea
\widetilde V_+&=&-\Bigl[(\CJ_0-\alpha)P_+
+\bigl(P_t-E\bigr)\CJ_+\Bigr]\,,
\label{V+til}\\[4pt]
\widetilde V_-&=&(\CJ_0+\alpha)P_-
+\frac{1}{c^{2}}\,\bigl(P_t+E\bigr)\,\CJ_-\,,
\label{V-til}\\[4pt]
\widetilde V_0&=&\frac1c\Bigl[-\alpha\bigl(P_t-E\bigr)
+E\,\N-\frac12 P_+\CJ_-\Bigr]
+\frac{c}{2}\,P_-\CJ_+\,.
\label{V0til}
\eea
In component form, the transformed system
$\widetilde V_\pm\Phi=0$ reads
\bea
&\sqrt{n+1}\,P_+\phi_{n+1}+\sqrt{n+2\alpha}\,\bigl(P_t-E\bigr)\phi_n=0\,,&
\label{I-exact}\\[4pt]
&c^{2}\,\sqrt{n+2\alpha}\,P_-\phi_n
+\sqrt{n+1}\;\bigl(P_t+E\bigr)\,\phi_{n+1}=0\,.&
\label{II-exact}
\eea
So far we have merely rewritten the relativistic theory; no limit has
been taken.

Two comments on \eqref{V+til}--\eqref{II-exact} are in order.
First, $\widetilde V_+$ --- equivalently Eq.~\eqref{I-exact} --- is
independent of $c$ (with fixed $E$), and the whole $c$-dependence of the
component system \eqref{I-exact}, \eqref{II-exact} is confined to the
single factor $c^{2}$ in \eqref{II-exact}.
Equation \eqref{I-exact} is \emph{chiral} in the sense
that it involves the momenta only through the single complex combination
$P_+=-2i\,\partial/\partial x_-$, the derivative with respect to one of
the two complex coordinates $x_\pm$; this chiral half of the system of
equations survives untouched in the singular limit, and the Carrollian
dynamics is decided entirely by what remains of \eqref{II-exact}.
Second, the two operators $P_t\mp E$ appearing in
\eqref{I-exact}, \eqref{II-exact} refer the frequency to the two
relativistic rest-energy branches $cP^{0}=\pm mc^{2}$: neither branch is
privileged, and the Carrollian spectrum will come out symmetric about
zero.  This symmetric placement reflects the phase-free choice of energy
origin.  An overall time-dependent redefinition
$\Phi_\delta=e^{i\delta E t}\Phi$, obtained from \eqref{SC} by $S \to e^{i\delta E t}S$, replaces $P_t$ by $P_t+\delta E$ in the
wave operators and shifts the two frequencies according to
\[
\{-E,E\}\longrightarrow\{-E-\delta E,E-\delta E\}.
\]
Their separation remains $2E$, and the common central shift does not
alter the Carroll covariance of the system. The choice $\delta E=E$,
which gives the pair $\{-2E,0\}$, is the Galilean phase-dressed frame
used in Ref.~\cite{Horvathy:2010vm}, as will be seen in
Section~\ref{sec:branching}.

\subsection{The Carrollian anyon wave equations}

Taking the fixed-$E$ Carroll limit \eqref{Clim} of the exact
system \eqref{V+til}--\eqref{V0til}, we obtain the Carrollian
anyon wave equations
\be
\label{CarrollEq}
\mathfrak V^{C}_0\Phi=0\,,\qquad
\mathfrak V^{C}_+\Phi=0\,,\qquad
\mathfrak V^{C}_-\Phi=0\,,
\ee
where
\bea
\mathfrak V^{C}_+&:=&-\widetilde V_+
=(\CJ_0-\alpha)P_++\bigl(P_t-E\bigr)\CJ_+\,,
\label{VC+}\\[4pt]
\mathfrak V^{C}_-&:=&
\lim_{\substack{c\to0\\\text{fixed-}E}}
c^{2}
\,\widetilde V_-
=\bigl(P_t+E\bigr)\;\CJ_-\,,
\label{VC-}\\[4pt]
\mathfrak V^{C}_0&:=&\lim_{\substack{c\to0\\\text{fixed-}E}}
c\,\widetilde V_0
=-\alpha\bigl(P_t-E\bigr)
+E\,\N-\frac12 P_+\CJ_-\,.
\label{VC0}
\eea
Let us analyse the consistency conditions that follow from the constraint
algebra:
\bea
[\mathfrak V^C_+,\mathfrak V^C_-]
&=&-P_+\mathfrak V^C_--2\mathcal D_C\CJ_0\,,
\label{Vcomm+-}\\[4pt]
[\mathfrak V^C_0,\mathfrak V^C_-]
&=&-E\mathfrak V^C_-\,,
\label{Vcomm0-}\\[4pt]
[\mathfrak V^C_0,\mathfrak V^C_+]
&=&P_+\mathfrak V^C_0-P_t\mathfrak V^C_+
+\mathcal D_C\CJ_+\,,
\label{Vcomm0+}
\eea
together with
\be
[\mathcal D_C,\mathfrak V^C_a]=0\,,\qquad [\mathcal D_C,\CJ_a]=0\,,\qquad a=0,\pm\,.
\label{DVcomm}
\ee
Thus $\mathcal D_C$, defined in \eqref{CarrollCasEq}, is the only secondary (central) operator generated by
the commutators, and closure of the vector constraints implies
$\mathcal D_C=0$, precisely the Carroll
equation \eqref{CarrollCasEq}.
Indeed, on the common kernel of \eqref{CarrollEq}, equations
\eqref{Vcomm+-} and \eqref{Vcomm0+} imply, respectively,
$\mathcal D_C\CJ_0\Phi=\mathcal D_C\CJ_+\Phi=0$.  In every non-trivial
irreducible module considered here, the common kernel of $\CJ_0$ and
$\CJ_+$ is $\{0\}$: $\CJ_+$ has no kernel in an infinite-dimensional
lowest-weight module, while in a finite spin-$j>0$ module its kernel is
the highest-weight state, on which $\CJ_0$ has eigenvalue $j$.

It is useful to verify this result componentwise, both as a check and as
preparation for constructing the solutions.  Using \eqref{Dplus}, the
$\mathfrak V^C_0$ equation becomes
\be
\label{V0comp}
\bigl(\mathfrak V^{C}_0\Phi\bigr)_n
=-\alpha\bigl(P_t-E\bigr)\phi_n
+E\,n\,\phi_n
-\frac12\,C^{\alpha}_{n}\,P_+\phi_{n+1}\,.
\ee
whereas the $\mathfrak V^C_\pm$ equations read
\bea
&\sqrt{n+2\alpha}\,\bigl(P_t-E\bigr)\phi_n
+\sqrt{n+1}\,P_+\phi_{n+1}=0\,,\qquad n\geq0\,,&
\label{C1}\\[4pt]
&\bigl(P_t+E\bigr)\,\phi_{n}=0\,,\qquad n\geq1\,.&
\label{C2}
\eea
The latter two equations are equivalent to \eqref{I-exact} and
\eqref{II-exact} with $c=0$.
Here $n$ ranges up to $\infty$ for anyons ($\alpha>0$, or
$0>\alpha\neq-j$) and up to $2j$ for bosons/fermions ($\alpha=-j$),
provided $j$ is a positive half-integer or integer.
Equation \eqref{C2} fixes every higher component of the tower to the
frequency $-E$ and therefore gives
$\mathcal D_C\phi_n=0$ for $n\geq1$.

At the lowest level, acting with $P_t+E$ on \eqref{C1} at $n=0$
and using \eqref{C2} at $n=1$ gives
$\sqrt{2\alpha}\,\mathcal D_C\phi_0=0$.  Hence
$\mathcal D_C\phi_0=0$ in every non-trivial module ($\alpha\neq0$).
Unlike the higher components, $\phi_0$ may contain modes on both
branches $P_t=\pm E$.  Thus every component satisfies the Carroll
equation \eqref{CarrollCasEq}.  Using \eqref{C2} in \eqref{C1}
for $n\geq1$ gives the recursion relation
\begin{equation}\label{recursion}
P_+\phi_{n+1}=2E\sqrt{\tfrac{n+2\alpha}{n+1}}\,\phi_n,\qquad n\geq1,
\end{equation}
on the negative energy branch,  and the positive energy modes are absent.
In terms of the original relativistic energy,
$cP^{0}=\pm\sqrt{m^{2}c^{4}+c^{2}\vec p^{\,\,2}}$, the fixed-$E$
Carroll limit retains both rest energies $\pm mc^{2}$ (with infinite $m$).  This
contrasts with the Galilean limit, which selects a single branch, as
shown in Section~\ref{sec:branching}.

Observe that multiplying \eqref{C1} by
$-\tfrac12\sqrt{n+2\alpha}$ and adding $\tfrac n2$ times
\eqref{C2} reproduces \eqref{V0comp}.  For $n=0$, the contribution
from \eqref{C2} vanishes identically, so \eqref{C1} alone suffices.
Thus, the three pieces of the Carrollian system \eqref{CarrollEq} are
complementary, but the pair $\mathfrak V^C_\pm\Phi=0$ already determines
the Carrollian wave equation completely.
The system \eqref{C1}, \eqref{C2} is therefore a first-order (in time and space)
\emph{square root} of the Carroll equation \eqref{CarrollCasEq}, in the same
sense in which the L\'evy-Leblond system is a square root of the
Schr\"odinger equation.

\subsection{Solutions}
\label{sec:solutions}

Equation \eqref{C2} makes all higher components monochromatic, of frequency $-E$,
\be
\label{sol1}
\phi_n(t,\vec x)=e^{+iEt}\,\varphi_n(\vec x)\,,\qquad n\geq1\,,
\ee
while \eqref{C1} with $n\geq1$ becomes the algebraic recursion for these negative energy profiles,
\be
\label{recur}
\varphi_n=\frac{1}{2E}\,\sqrt{\frac{n+1}{n+2\alpha}}\;
P_+\varphi_{n+1}\,,
\qquad n\geq1\,,
\ee
or, read upwards in momentum space $P_+\to p_+=p_1+ip_2$, with $(a)_k=a(a+1)\cdots(a+k-1)=\Gamma(a+k)/\Gamma(a)$ the Pochhammer symbol,
\be
\label{recurup}
\varphi_{n}=\Bigl(\frac{2E}{p_+}\Bigr)^{n-1}
\sqrt{\frac{(1+2\alpha)_{n-1}}{n!}}\;\varphi_1\,,\qquad n\geq1\,,
\ee
which presupposes $p_+\neq0$. The point $\vec p=0$ is treated separately in
remark \textbf{(d)} below. Finally, \eqref{C1} with $n=0$
determines the time dependence of the lowest component,
\be
\label{sol0}
\phi_0(t,\vec x)=e^{-iEt}\,\chi(\vec x)
+e^{+iEt}\,\frac{1}{2E\sqrt{2\alpha}}\,P_+\varphi_1(\vec x)\,,
\ee
with $\chi(\vec x)$ arbitrary.  Hence the general solution contains
\emph{both} Carroll sectors simultaneously: a positive-energy sector at
$+E$, carried by the arbitrary function $\chi(\vec x)$ in the lowest
component, and a negative-energy sector at $-E$, which spans the whole
spin tower and is generated from the single
datum $\varphi_1(\vec x)$.

Five remarks are in order.
\begin{enumerate}

\item[\textbf{(a)}] \textit{Termination for (half-)integer spin.}
If $2\alpha=-2j$ is a
negative integer, $(1+2\alpha)_{n-1}=(1-2j)_{n-1}$ vanishes for
$n\geq2j+1$ in \eqref{recurup}, so the tower terminates automatically at $n=2j$, as it must
for a $(2j+1)$-component field.  The top component $\varphi_{2j}$ is then
the free datum of the negative-energy sector, with all lower components
determined by \eqref{recur}; explicit examples are given in
Section~\ref{sec:examples}.

\item[\textbf{(b)}] \textit{Internal-module convergence at fixed spin.}

For the infinite-dimensional representations, \eqref{recurup} gives
\[
\left|\frac{\varphi_{n+1}}{\varphi_n}\right|
   \longrightarrow \frac{2E}{|\vec p\,|}\,.
\]
In the units inherited from the relativistic parent theory, $E$ and
$|\vec p\,|$ have the same dimensions, so this ratio is dimensionless.
By the ratio test the
$\so(2,1)$-module norm $\sum_n|\varphi_n|^{2}$ converges if and only if
$|\vec p\,|>2E$, which is therefore the normalisability condition.  At the boundary
$|\vec p\,|=2E$, one has
\be
|\varphi_n|^2
=\frac{\Gamma(n+2\alpha)}
{\Gamma(1+2\alpha)\Gamma(n+1)}\,|\varphi_1|^2
\sim
\frac{n^{2\alpha-1}}{\Gamma(1+2\alpha)}\,|\varphi_1|^2\,.
\ee
Consequently, the boundary norm behaves as $\sum_n n^{2\alpha-1}$
and converges only for $\alpha<0$.  Within the first half-bounded
non-unitary interpolation sector, where these component norms remain
real, this is precisely the window $-\tfrac12<\alpha<0$; for
$\alpha=0$ the norm is harmonic and diverges, while for $\alpha>0$ it
diverges by a power.  For finite spin the question does not arise.

\item[\textbf{(c)}] \textit{Anyon interpolation.}
In the interval $-j<\alpha<-j+\frac12$ the
non-unitary representations $\tilde D^{+}_\alpha$ interpolate between the
Carrollian spin-$(j-\frac12)$ and spin-$j$ systems, exactly as in the
relativistic and Galilean cases \cite{Horvathy:2010vm}, the components with
$n>2j$ being suppressed by the additional factor $(j+\alpha)^{1/2}$.

\item[\textbf{(d)}] \textit{The zero-momentum sector.}
The inversion \eqref{recurup} is not
available at $p_+=0$, so this point requires a direct analysis of
\eqref{C1}, \eqref{C2}.  At zero spatial momentum the recursion term drops
from \eqref{C1}, which reduces to
$\sqrt{n+2\alpha}\,\bigl(P_t-E\bigr)\phi_n=0$ at every
level.  For a genuine anyon the coefficient never vanishes, and combining
with the branch condition \eqref{C2} gives $2E\,\phi_n=0$ for
$n\geq1$: the tower sector disappears entirely, and the general solution
reduces to $\phi_0=e^{-iEt}\chi$.  The zero-momentum Carrollian
anyon therefore occupies the lowest component only, on the
positive-energy branch $+E$.  This is consistent with remark
\textbf{(b)}: the tower
sector requires $|\vec p\,|>2E$, so the point $\vec p=0$ was already
excluded from its support, whereas the positive-energy sector exists for
every momentum.  For finite spin $\alpha=-j$ the top component
escapes the argument, since its equation in \eqref{C1} carries a vanishing
coefficient, \cf\ remark \textbf{(a)}; hence at $\vec p=0$ two data survive
--- the top component $\varphi_{2j}$ on the $-E$ branch and the
positive-energy mode
$\chi$ in $\phi_0$ --- with all intermediate components vanishing, as is
immediate for the Carrollian L\'evy-Leblond system \eqref{CLL} below. Finally, note that $\vec p=0$ is not a Carroll-invariant locus:
since
\[
[B_i,P_j]=i\delta_{ij}P_t
\]
and $P_t=\pm E\neq0$ on both branches, a Carroll boost shifts
the momentum.  Hence the zero-momentum state belongs to the same
Carroll orbit as the arbitrary-momentum positive-energy modes carried
by the datum $\chi$ in \eqref{sol0}.  This is a transformation between
states, not spatial propagation: the Carroll dispersion relation is
independent of $\vec p$.

\item[\textbf{(e)}] \textit{Holomorphic zero modes.}
Remark \textbf{(d)} generalises.  Since
$P_+=-2i\,\partial/\partial x_-$, the kernel of $P_+$ consists of the
\emph{holomorphic} functions $\phi(x_+)$.  Among tempered distributions
these are the polynomials in $x_+$, whose Fourier transforms are supported
at the single point $\vec p=0$: the holomorphic sector is the
zero-momentum sector in the generalised (distributional) sense, and the
analysis of \textbf{(d)} applies to it verbatim.  (Entire non-polynomial
elements of the kernel, such as $e^{\kappa x_+}$, correspond to complex
momenta, $p_+=0$, $p_-=-2i\kappa$.)  These zero modes have a structural
consequence.  The recursion \eqref{recur} determines each component
from the one \emph{above} it, so reconstructing the tower upwards, as in
\eqref{recurup}, is ambiguous precisely by an element of $\ker P_+$ at
each level: inserting a holomorphic $h(x_+)$ at level $N$ leaves all lower components untouched and dresses the higher ones by terms
$\propto x_-^{\,k}\,h(x_+)$.
To see explicitly, set
\be
b_n:=2E\sqrt{\frac{n+2\alpha}{n+1}}
\ee
and add $\delta\varphi_N=h_N(x_+)\in\ker P_+$.  The downward recursion
then gives
\be
 \delta\varphi_{N-1}=b_{N-1}^{-1}P_+h_N=0
 \qquad\Longrightarrow\qquad
 \delta\varphi_n=0\quad \text{for} \quad n<N,
\ee
whereas the upward equations require
\bea
 P_+\delta\varphi_{N+k+1}&=&b_{N+k}\delta\varphi_{N+k},
 \nonumber\\
 \delta\varphi_{N+k}&=&
 \frac{(iE)^k}{k!}
 \sqrt{\frac{(N+2\alpha)_k}{(N+1)_k}}\,
 x_-^k h_N(x_+),\qquad k\geq0.
\eea

Thus the lower components remain unchanged, but the required upper tail
changes the complete tower.  In a finite-dimensional module
$\alpha=-j$ the formula stops at $k=2j-N$ and changes the top component
$\varphi_{2j}$.  The tower sector therefore carries \emph{holomorphic
moduli}, with no counterpart in the Galilean mirror of
Ref.~\cite{Horvathy:2010vm}; different choices of $h_N$ determine
distinct complete solutions.

The holomorphic deformation remains convergent in the internal
spin-tower direction.  Indeed,
\be
 \left|
 \frac{\delta\varphi_{N+k+1}}
      {\delta\varphi_{N+k}}
 \right|^2
 =
 \frac{E^2|x_-|^2}{(k+1)^2}
 \left|\frac{N+2\alpha+k}{N+1+k}\right|
 \longrightarrow0 ,
\ee
so the corresponding series has infinite radius of convergence for
every finite $x_-$.  It is therefore a genuine solution of the complete
wave system, rather than a formal failure of the upward recursion.

The holomorphic moduli are excluded only if the physical solution space
is restricted to ordinary square-integrable wave packets with momentum
support in $|\vec p\,|>2E$.  Their Fourier transforms are distributions
supported at $p_+=0$, equivalently $\vec p=0$, and their norm on the free
spatial plane diverges.  Within that restricted wave-packet sector,
$P_+$ is invertible and the negative-energy tower is completely
parametrised by $\varphi_1$.

\end{enumerate}

\section{Carroll symmetry}
\label{sec:carrollalgebra}

Starting from the Poincar\'e generators \eqref{Pgen} and their algebra
\eqref{PoincareAlg}, the rotation is generated by
$\mathcal M_0=\epsilon_{ij}x_iP_j+\CJ_0$, while
$\mathcal M_\pm=\pm i(x^{0}P_\pm-x_\pm P^{0})+\CJ_\pm$ generate the
Lorentz boosts.  Introduce the combinations
$K_i=-c^{-1}\epsilon_{ij}\mathcal M_j$.  In the complex basis of
\eqref{xPcomm}, using $P^0=P_t/c$ from \eqref{Pgen}, they read
\be
\label{Kpm}
K_\pm=K_1\pm iK_2
=\pm\frac{i}{c}\,\mathcal M_\pm
=-tP_\pm+\frac{x_\pm}{c^{2}}P_t
\pm\frac{i}{c}\,\CJ_\pm\,.
\ee
Applying the grading \eqref{SC} and using \eqref{tilCJ} gives the exact
operators
\be
\label{Ktil}
\widetilde K_+=-tP_++\frac{x_+}{c^{2}}P_t+i\,\CJ_+\,,
\qquad
\widetilde K_-=-tP_-+\frac{x_-}{c^{2}}P_t-\frac{i}{c^{2}}\,\CJ_-\,.
\ee
The fixed-$E$ Carroll limit \eqref{Clim} therefore yields
\be
\label{Bpm}
B_\pm:=\lim_{c\to0}c^{2}\widetilde K_\pm\,,\qquad
B_+=x_+P_t\,,\qquad
B_-=x_-P_t-i\CJ_-\,.
\ee
The rotation generator is unaffected by the grading and by the limit,
\be
\label{Jrot}
\mathrm J=\epsilon_{ij}x_iP_j+\CJ_0\,.
\ee
Using \eqref{xPcomm}, one verifies that
$\{P_t,P_i,B_i,\mathrm J\}$ close on the $(2+1)$-dimensional Carroll
algebra,
\bea
&[B_i,P_j]=i\delta_{ij}P_t\,,\qquad
[P_i,P_j]=0\,,\qquad
[B_i,B_j]=0\,,\qquad
[P_t,P_i]=[P_t,B_i]=0\,,&
\label{Car1}\\[4pt]
&[\mathrm J,P_i]=i\epsilon_{ij}P_j\,,\qquad
[\mathrm J,B_i]=i\epsilon_{ij}B_j\,,\qquad
[\mathrm J,P_t]=0\,.&
\label{Car2}
\eea
Their action on the wave operators \eqref{VC+}--\eqref{VC0} makes the
covariance of the system equally explicit.  The commutators that do not
vanish identically close on the wave operators themselves,
\bea
[\mathrm J,\mathfrak V^C_\pm]&=&\pm\mathfrak V^C_\pm\,,
\nonumber\\[3pt]
[B_-,\mathfrak V^C_+]&=&2i\mathfrak V^C_0\,,\qquad
[B_-,\mathfrak V^C_0]=-i\mathfrak V^C_-\,.
\label{CarrollWaveCov}
\eea
All the remaining commutators vanish identically:
$[\mathrm J,\mathfrak V^C_0]=[B_-,\mathfrak V^C_-]=0$,
$[B_+,\mathfrak V^C_a]=0$, and
$[P_t,\mathfrak V^C_a]=[P_i,\mathfrak V^C_a]=0$, with $a=0,\pm$.
Consequently, for any
$G\in\{P_t,P_i,B_\pm,\mathrm J\}$ and every solution of
\eqref{CarrollEq},
\be
\mathfrak V^C_a(G\Phi)
=G\bigl(\mathfrak V^C_a\Phi\bigr)
-[G,\mathfrak V^C_a]\Phi=0\,.
\label{CarrollWaveSym}
\ee
Thus the Carroll generators preserve the common kernel of the wave
operators and are genuine on-shell symmetries of the first-order system.

Note that the time-translation operator $P_t$ introduced in \eqref{Pgen}
takes over the role played by the central mass in the Bargmann algebra.
Unlike the Bargmann mass, however, it is both a central charge and the
Hamiltonian itself.

\subsection{Casimir operators and Carrollian spin}

The two Carroll invariants, and the values they take,
emerge directly as $c\to0$ limits of the relativistic Poincar\'e
Casimirs \eqref{PCas}.  In components, $\mathcal C_1=-(P^{0})^{2}+P_+P_-$
and, using $P_1\CJ_1+P_2\CJ_2=\tfrac12(P_+\CJ_-+P_-\CJ_+)$,
$\mathcal C_2=P^{0}\CJ_0+\tfrac12(P_+\CJ_-+P_-\CJ_+)$.  Employing the similarity transformation
\eqref{tilCJ} we get
\be
\label{tilPCas}
S\,\mathcal C_1\,S^{-1}=-\frac{P_t^{2}}{c^{2}}+P_+P_-\,,
\qquad
S\,\mathcal C_2\,S^{-1}
=\frac{1}{c}\Bigl(P_t\CJ_0+\tfrac12P_+\CJ_-\Bigr)
+\frac{c}{2}\,P_-\CJ_+\,.
\ee
Multiplying by the powers of $c$ that cancel the leading divergences and
letting $c\to0$ at fixed $E=mc^2$ gives
\be
\label{PCaslim}
\widehat{\mathcal{C}}_1:=\lim_{c\to0}c^{2}
\bigl(S\mathcal C_1S^{-1}\bigr)=-P_t^{2}\,,
\qquad
\widehat{\mathcal{C}}_2:=\lim_{c\to0}c\bigl(S\mathcal C_2S^{-1}\bigr)
=P_t\CJ_0+\tfrac12P_+\CJ_-\,.
\ee
Being $c\to0$ limits of Poincar\'e invariants, $\widehat{\mathcal C}_1$
and $\widehat{\mathcal C}_2$ are themselves invariants of the Carroll
algebra \eqref{Car1}--\eqref{Car2}.  Their on-shell values follow by
contracting the Klein--Gordon and Pauli--Lubanski conditions
\eqref{KGPL}.  Indeed, $(\mathcal C_1+m^{2}c^{2})\psi=0$ becomes, after
conjugating by $S$ and multiplying by $c^{2}$,
\be
\label{C1lim}
\Bigl(-P_t^{2}+c^{2}P_+P_-+m^{2}c^{4}\Bigr)\Phi=0\,.
\ee
Taking $c\to0$ at fixed $E=mc^{2}$, the middle term drops and
$m^{2}c^{4}\to E^{2}$, so \eqref{C1lim} becomes
$\mathcal D_C\Phi=0$.  Thus the
$c\to0$ limit of the Klein--Gordon equation reproduces the Carroll
equation \eqref{CarrollCasEq}, in precise analogy with the Galilean $c\to\infty$ limit of the
same relativistic condition, which instead produces the free
Schr\"odinger equation.

For the Pauli--Lubanski condition, $(\mathcal C_2-\alpha mc)\psi=0$, conjugating by $S$ and multiplying by $c$
gives $c\,\bigl(S\mathcal C_2S^{-1}\bigr)\Phi=\alpha mc^{2}\Phi$, and
since $mc^{2}=E$ is held fixed, the
right-hand side is already $c$-independent.  Taking $c\to0$ on the left
with \eqref{PCaslim} gives directly
$\widehat{\mathcal C}_2\Phi=E\,\alpha\,\Phi$.
Collecting both results, the Carroll invariants \eqref{PCaslim} take
the form
\be
\label{CarCas}
\widehat{\mathcal C}_1 \Phi
=-P_t^{2}\,,
\qquad
\widehat{\mathcal C}_2
=P_t\mathrm J-\epsilon_{ij}B_iP_j\,,
\ee
with eigenvalues
\be
\label{CasValues}
\widehat{\mathcal C}_1\Phi=-E^2\Phi\,,\qquad
\widehat{\mathcal C}_2\Phi=E\,\alpha\Phi\,.
\ee
The first eigenvalue condition is equivalent to the Carroll equation \eqref{CarrollCasEq},
$(P_t^2-E^2)\Phi=0$.
$\widehat{\mathcal C}_2$, by contrast, gives the independent label of
Carrollian spin.  The system \eqref{C1}, \eqref{C2} describes a
Carroll particle with energy gap $2E$ and spin $\alpha$.

\subsection{Classification of the Carrollian sectors}
\label{sec:classes}

The grading exponent of the similarity transformation \eqref{SC} is a free parameter.
Let us modify \eqref{SC}, introducing a new $\beta$-parameter,
\be
\label{Mbeta}
S_\beta=c^{\,\beta\N}\,,
\qquad \phi_n=c^{\,\beta \,n}\psi_n\,,
\qquad
\widetilde\CJ_\pm=c^{\pm\beta}\CJ_\pm\,.
\ee

The exact component
equations become
\bea
&\sqrt{n+2\alpha}\,(P_t-E)\phi_n
+c^{1-\beta}\,\sqrt{n+1}\,P_+\phi_{n+1}=0\,,&
\label{Ibeta}\\[4pt]
&c^{1+\beta}\,\sqrt{n+2\alpha}\,P_-\phi_n
+\sqrt{n+1}\,(P_t+E)\phi_{n+1}=0\,.&
\label{IIbeta}
\eea
For $c\to0$ there are two non-degenerate classes, the endpoints
$\beta=\pm1$.

\begin{itemize}

\item For $\beta=+1$ both terms of \eqref{Ibeta} survive and
\eqref{IIbeta} reduces to the (negative energy) branch condition: this is the system of
Section~\ref{sec:carroll}, with the spin block in the Carroll boost
$B_-$ \eqref{Bpm} and the two
energy branches coexisting.

\item For $\beta=-1$ the roles are exchanged: \eqref{Ibeta} degenerates to
$(P_t-E)\phi_n=0$ for every $n$, and the system reads
\be
\label{dual}
(P_t-E)\phi_n=0\,,\qquad
\sqrt{n+2\alpha}\,P_-\phi_n+2E\,\sqrt{n+1}\,\phi_{n+1}=0\,,
\ee
where the second equation used the first at level $n+1$,
$(P_t+E)\phi_{n+1}=2E\,\phi_{n+1}$: the whole tower sits on the
positive-energy branch and is generated
upwards from $\phi_0$ by $P_-/2E$.  Here the redefinition
$\phi_n\to e^{-iEt}\,(2E)^{-n}\phi_n$ removes $E$
altogether --- the energy parameter is \emph{removable}, in sharp contrast
with $\beta=+1$, and the system becomes genuinely static. This is the single-frequency, phase-removable endpoint.  After removing
the common phase, the tower is strictly time independent, although its
non-trivial spatial recursion through $P_-$ survives.

\item For the intermediate central class
$-1<\beta<1$, represented by the element $\beta=0$, both recursion terms vanish, so
$(P_t-E)\phi_n=0$ for all $n$ and $(P_t+E)\phi_n=0$ for
$n\geq1$: only
$\phi_0=e^{-iEt}\chi(\vec x)$ survives and the boosts become
purely orbital; the degenerate ``central'' class, in which the spin
content is lost.

\item For arbitrary $\beta$, the transformed relativistic boosts are
\be
\label{KtilBeta}
\widetilde K^{(\beta)}_+
=-tP_++\frac{x_+}{c^2}P_t+i\,c^{\beta-1}\CJ_+\,,
\qquad
\widetilde K^{(\beta)}_-
=-tP_-+\frac{x_-}{c^2}P_t-i\,c^{-\beta-1}\CJ_-\,.
\ee
The Carroll normalisation $c^2\widetilde K_i^{(\beta)}$ is the unique
common scaling that retains both the orbital term $x_iP_t$ and the
non-zero bracket $[B_i,P_j]=i\delta_{ij}P_t$.  It gives
\be
c^2\widetilde K^{(\beta)}_+
\longrightarrow x_+P_t+i\,c^{\beta+1}\CJ_+\,,
\qquad
c^2\widetilde K^{(\beta)}_-
\longrightarrow x_-P_t-i\,c^{1-\beta}\CJ_-\,.
\ee
Finiteness requires simultaneously $\beta+1\geq0$ and
$1-\beta\geq0$, hence $-1\leq\beta\leq1$.  For $\beta>1$ the spin
part of $\widetilde K_-^{(\beta)}$ diverges, whereas for $\beta<-1$
the spin part of $\widetilde K_+^{(\beta)}$ diverges.  An additional
rescaling could suppress that divergence only by eliminating the
orbital Carroll boost and degenerating the boost--translation bracket;
it would therefore define a different, degenerate contraction.  This
reproduces, for arbitrary spin, exactly the three boost classes
$\beta_\star=+1,0,-1$ and the exclusion of $|\beta|>1$ found for the Dirac
field in \cite{Parekh:2026wri}.
\end{itemize}

\section{Carrollian bosons, fermions and higher-spin fields}
\label{sec:examples}

Throughout this section we specialise to the non-trivial
finite-dimensional modules $\alpha=-j$, with
$j=\tfrac12,1,\tfrac32,\ldots$, and use the conventions \eqref{C'D} of
Appendix~\ref{app:conv}.  The spin tower then truncates to
$\Phi=(\phi_0,\ldots,\phi_{2j})^T$.  Since these systems are finite
truncations of \eqref{CarrollEq}, the constraint-algebra argument
\eqref{Vcomm+-}--\eqref{DVcomm} applies uniformly: every component
satisfies the Carroll equation \eqref{CarrollCasEq}.

\subsection{Spin \texorpdfstring{$1/2$}{1/2}: the Carrollian L\'evy-Leblond system}

For $\alpha=-j=-\tfrac12$, $\Phi=(\phi_0,\phi_1)^T$, and
Eqs.~\eqref{C1}, \eqref{C2} reduce to
\be
\label{CLL}
\bigl(P_t-E\bigr)\phi_0+P_+\phi_1=0\,,
\qquad
\bigl(P_t+E\bigr)\,\phi_1=0\,.
\ee
Acting with $P_t+E$ on the first equation in \eqref{CLL} and using
the second gives $\mathcal D_C\phi_0=0$, while acting with $P_t-E$ on
the second gives $\mathcal D_C\phi_1=0$. Thus all components satisfy the Carroll equation \eqref{CarrollCasEq}. We therefore refer to
\eqref{CLL} as the \emph{Carrollian L\'evy--Leblond system}, by analogy
with the Galilean L\'evy--Leblond equations, which yield the
Schr\"odinger equation as their integrability condition
\cite{Levy-Leblond:1967eic,Horvathy:2010vm}.  Its general solution is
\be
\phi_0=e^{-iEt}\chi(\vec x)
+e^{+iEt}\,\frac{1}{2E}\,P_+\varphi_1(\vec x)\,,\qquad\phi_1=e^{+iEt}\varphi_1(\vec x)\,,
\ee
with $\chi,\varphi_1$ arbitrary.

The Carroll boosts \eqref{Bpm}, derived in
Section~\ref{sec:carrollalgebra}, read in matrix form on
$\Phi=(\phi_0,\phi_1)^{T}$
\be
\label{CLLboost}
B_+=x_+\,P_t\,,\qquad
B_-=x_-\,P_t+
\begin{pmatrix} 0 & i\\ 0&0\end{pmatrix}\,,
\qquad
\mathrm J=\epsilon_{ij}x_iP_j+
\begin{pmatrix} -\tfrac12 & 0\\ 0&\tfrac12\end{pmatrix}\,.
\ee
The structure ``diagonal
orbital piece $x_i\,P_t$ plus a single nilpotent off-diagonal spin block'' is also observed in $3+1$ dimensional spin-$\tfrac 12$ fields \cite{Parekh:2026wri}.

\subsection{Spin 1}

For $\alpha=-j=-1$, $\Phi=(\phi_0,\phi_1,\phi_2)^T$, and
Eqs.~\eqref{C1}, \eqref{C2} give
\be
\label{CDJT}
\sqrt2\,\bigl(P_t-E\bigr)\phi_0+P_+\phi_1=0\,,\qquad
\bigl(P_t-E\bigr)\phi_1+\sqrt2\,P_+\phi_2=0\,,\qquad
\bigl(P_t+E\bigr)\phi_1=\bigl(P_t+E\bigr)\phi_2=0\,.
\ee
The Carroll equation follows component by component.  The last two
relations in \eqref{CDJT} immediately give
\[
\mathcal D_C\phi_1=\mathcal D_C\phi_2=0\,.
\]
Acting with $P_t+E$ on the first relation and using
$(P_t+E)\phi_1=0$ gives
\[
\sqrt2\,\mathcal D_C\phi_0+P_+(P_t+E)\phi_1
=\sqrt2\,\mathcal D_C\phi_0=0\,.
\]
Thus $\mathcal D_C\phi_n=0$ for $n=0,1,2$.
The general solution of \eqref{CDJT} is
\bea
\phi_2=e^{+iEt}\varphi_2(\vec x)\,,\qquad
\phi_1=e^{+iEt}\,\frac{\sqrt2}{2E}\,P_+\varphi_2\,,\qquad
\phi_0=e^{-iEt}\chi(\vec x)
+e^{+iEt}\,\frac{1}{(2E)^{2}}\,P_+^{2}\varphi_2\,,
\eea
where $\varphi_2$ is arbitrary.
This can be regarded as the Carroll limit of the Deser--Jackiw--Templeton
(topologically massive gauge field) system \cite{Deser:1981wh}.
Up to linear combinations, the three fields correspond to the components
of the \emph{dual field strength}
$\widetilde F_\mu=\frac12\epsilon_{\mu\nu\lambda}F^{\nu\lambda}$ of the
topologically massive gauge field.  The spin contribution to the boost
$B_-$ \eqref{Bpm} and the
rotation are
\be
\CJ_-=
\begin{pmatrix} 0& -\sqrt2&0\\ 0&0&-\sqrt2\\ 0&0&0\end{pmatrix},
\qquad
\CJ_0=\mathrm{diag}(-1,0,1)\,.
\ee

\subsection{Spin-\texorpdfstring{$\frac{3}2$}{3/2} and higher-spin Carroll fields}
\label{sec:higherspin}

The pattern of the two examples above extends verbatim to every
(half-)integer $j$, providing free \emph{higher-spin Carroll fields}.
For general $j$, Eqs.~\eqref{C1}, \eqref{C2} take the universal form
\be
\label{Chs}
\sqrt{2j-n}\,\bigl(P_t-E\bigr)\phi_n
+\sqrt{n+1}\,P_+\phi_{n+1}=0\,,\quad n=0,\dots,2j-1\,,
\qquad
\bigl(P_t+E\bigr)\phi_n=0\,,\quad n\geq1\,,
\ee
whose general solution takes the form
\be
\label{hssol}
\phi_n=e^{-iEt}\,\chi(\vec x)\,\delta_{n,0}
+e^{+iEt}\,\frac{1}{(2E)^{2j-n}}\sqrt{\binom{2j}{n}}\;
P_+^{\,2j-n}\,\varphi_{2j}(\vec x)\,,
\qquad n=0,\dots,2j\,,
\ee
as follows by iterating the recursion contained in \eqref{Chs}
(it reproduces the spin-$\tfrac 12$ and spin-$1$ solutions above). For
$\alpha=-\frac32$ it gives the four-component Carrollian
Rarita--Schwinger system
\bea
\sqrt3\,(P_t-E)\phi_0+P_+\phi_1&=&0\,,
\nonumber\\[3pt]
(P_t-E)\phi_1+P_+\phi_2&=&0\,,
\nonumber\\[3pt]
(P_t-E)\phi_2+\sqrt3\,P_+\phi_3&=&0\,,
\nonumber\\[3pt]
(P_t+E)\phi_1=(P_t+E)\phi_2=(P_t+E)\phi_3&=&0\,.
\label{CRaritaSchwinger}
\eea
The last line gives $\mathcal D_C\phi_n=0$ for $n=1,2,3$.
Acting with $P_t+E$ on the first line then yields
\[
\sqrt3\,\mathcal D_C\phi_0+P_+(P_t+E)\phi_1=0\,,
\]
and hence $\mathcal D_C\phi_0=0$.  Therefore all four components obey
the Carroll equation.

The structure of
\eqref{hssol} deserves comment.  The two physical data sit at the two
\emph{ends} of the spin tower: the top state $|2j)$ carries the free
datum $\varphi_{2j}$ of the negative-energy ($-E$) branch, and the
bottom state $|0)$ carries the
positive-energy mode $\chi$ on the $+E$ branch; every intermediate
component is a binomially weighted holomorphic descendant
$P_+^{\,2j-n}\varphi_{2j}$, with no independent content.  Unlike the
anyonic case, the downward recursion from the top requires no inversion
of $P_+$: the parametrisation \eqref{hssol} is exact and complete for
\emph{all} momenta --- the convergence condition of remark \textbf{(b)}
and the holomorphic moduli of remark \textbf{(e)} do not arise --- and at
$\vec p=0$ precisely the two endpoint data survive, in accordance with
remark \textbf{(d)}.

%%%%%%%%%%%%%%%%%%%%%%%%%%%%%%%%%%%%%%%%%%%%%%%%%%%%%%%%%%%%%%%%%%%%%%%%%%%%%%
\section{Exotic Carroll equation and exotic Carroll algebra}
\label{sec:exotic}
%%%%%%%%%%%%%%%%%%%%%%%%%%%%%%%%%%%%%%%%%%%%%%%%%%%%%%%%%%%%%%%%%%%%%%%%%%%%%%

\subsection{Exotic Carroll equation}

The exotic Galilean algebra is characterised by a non-vanishing boost
commutator, $[\CK_i,\CK_j]=-i\epsilon_{ij}\,\kappa_G$, where $\kappa_G$
is the exotic central charge.  At the Lie-algebra level, the planar
Carroll algebra likewise admits three independent central extensions
\cite{deAzcarraga:1997mdw,Matulich:2019cdo}; the one relevant here is the
exotic boost cocycle,
$[B_i,B_j]=-i\epsilon_{ij}\,\kappa_C$.
The relativistic boosts \eqref{Kpm} satisfy
$[K_i,K_j]=-\frac{i}{c^{2}}\epsilon_{ij}\,\mathcal M_0$, so that with the
Carrollian normalisation \eqref{Bpm},
\be
[B_i,B_j]=-i\epsilon_{ij}\lim_{c\to0}c^{2}\,\mathcal M_0\,.
\ee
A nonzero right-hand side therefore requires the rotation generator to increase as $c^{-2}$, that is, the spin itself must be scaled.  Since the spin part of the Lorentz generator is given by $\CJ_0=\N+\alpha$, such behaviour is achieved by letting the spin $\alpha$ itself diverge as $c^{-2}$. Define the
\emph{exotic Carroll limit}
\be
\label{exoC}
c\to0\,,\qquad \alpha\to\infty\,,\qquad
\kappa_C:=\alpha\,c^{2}=\text{fixed}\,,
\qquad E=mc^{2}=\text{fixed}\,,
\ee
which is the Carrollian mirror of the Jackiw--Nair scaling
$\alpha/c^{2}=\kappa_G$ of the exotic Galilean limit
\cite{Jackiw:2000tz,Duval:2000xr,Horvathy:2004fw}.

The contraction of the internal Lorentz algebra can be displayed
explicitly.  Recall from \eqref{SC} that $\N=\CJ_0-\alpha$.  At finite
$\alpha$ introduce the intermediate operators
\be
\widetilde b_\alpha^{\,\pm}:=\frac{\CJ_\pm}{\sqrt{2\alpha}}\,,\qquad
\I_\alpha:=\frac{\CJ_0}{\alpha}
=\I+\frac{\N}{\alpha}\,.
\label{FiniteOscillators}
\ee
At finite $\alpha$, the Lorentz commutators give
\be
[\widetilde b_\alpha^{-},\widetilde b_\alpha^{+}]=\I_\alpha\,,
\qquad
[\I_\alpha,\widetilde b_\alpha^{\,\pm}]
=\pm\frac{1}{\alpha}\widetilde b_\alpha^{\,\pm}\,,
\qquad
[\N,\widetilde b_\alpha^{\,\pm}]
=\pm\widetilde b_\alpha^{\,\pm}\,.
\label{FiniteOscillatorAlg}
\ee
Indeed, the first relation follows directly from
$[\CJ_-,\CJ_+]=2\CJ_0$.
On the basis \eqref{Dplus},
\bea
\widetilde b_\alpha^{+}|n)&=&
\sqrt{(n+1)\left(1+\frac{n}{2\alpha}\right)}\,|n+1)\,,
\nonumber\\[4pt]
\widetilde b_\alpha^{-}|n)&=&
\sqrt{n\left(1+\frac{n-1}{2\alpha}\right)}\,|n-1)\,,
\qquad
\I_\alpha|n)=\left(1+\frac{n}{\alpha}\right)|n)\,.
\label{FiniteOscillatorAction}
\eea
The limit exists on the dense algebraic span of the states $|n)$ with
fixed, finite $n$.  The contracted
operators themselves are given by
\bea
b^\pm&:=&\lim_{\alpha\to+\infty}
\widetilde b_\alpha^{\,\pm}
=\lim_{\alpha\to+\infty}\frac{\CJ_\pm}{\sqrt{2\alpha}}\,,
\qquad \lim_{\alpha\to+\infty}\I_\alpha=\I\,,
\nonumber\\[4pt]
[b^-,b^+]&=&\I\,,\qquad
[\N,b^\pm]=\pm b^\pm\,,\qquad
\N=b^+b^-\,.
\label{HeisContraction}
\eea
Thus, remarkably, the large-spin contraction turns the internal
$\so(2,1)$ algebra into the Heisenberg algebra, while the family of
discrete-series modules $D_\alpha^+$ contracts to the standard
oscillator Fock module.  The action of the contracted generators is
\bea
b^+|n)=\sqrt{n+1}\,|n+1)\,,
\qquad
b^-|n)=\sqrt n\,|n-1)\,,
\qquad \N|n)=n|n)\,.
\label{HeisContractionAction}
\eea
These are equalities in the limiting module.

Using $\alpha=\kappa_C/c^2$ in \eqref{exoC}, the scaled spin generators
obey
\be
\lim_{\substack{c\to0,\ \alpha\to+\infty\\
\alpha c^2=\kappa_C}}
c\CJ_\pm
=\lim_{\substack{c\to0,\ \alpha\to+\infty\\
\alpha c^2=\kappa_C}}
\sqrt{2\kappa_C}\,\widetilde b_{\alpha}^{\,\pm}
=\sqrt{2\kappa_C}\,b^\pm\,.
\label{scaledSpin}
\ee
We apply \eqref{exoC} directly to the Cort\'es--Plyushchay vector system
\eqref{intro:V}, in the component form \eqref{HP1}, \eqref{HP2}, without
a similarity transformation, \ie\ at the $\beta=0$ point of
\eqref{Mbeta}.  Multiplying \eqref{HP1} by $-c$ and \eqref{HP2} by $c$,
the exact equations read
\be
\label{exoexact}
\begin{aligned}
&\sqrt{n+2\alpha}\,\bigl(P_t-E\bigr)\psi_n
+c\sqrt{n+1}\,P_+\psi_{n+1}=0\,,
\\[4pt]
&c\sqrt{n+2\alpha}\,P_-\psi_n
+\sqrt{n+1}\,\bigl(P_t+E\bigr)\psi_{n+1}=0\,.
\end{aligned}
\ee
\ie\ \eqref{Ibeta}, \eqref{IIbeta} at $\beta=0$.  At fixed $\alpha$,
this point belongs to the degenerate central class of
Section~\ref{sec:classes}.  The simultaneous exotic scaling
$\alpha=\kappa_C/c^2$, however, compensates the vanishing powers of $c$
and restores a non-trivial component recursion.  Since
$\sqrt{n+2\alpha}=c^{-1}\sqrt{nc^{2}+2\kappa_C}$, the divergent
coefficient supplies one inverse power of $c$ exactly where needed: in
the first equation of \eqref{exoexact} (multiplied by $c$) the $P_+$
term is suppressed by $c^{2}$, while in the second the explicit factor
$c$ is cancelled and \emph{both} terms survive at order $c^{0}$.  The
limit is $(P_t-E)\psi_n=0$ together with
$\sqrt{2\kappa_C}\,P_-\psi_n+\sqrt{n+1}\,(P_t+E)\psi_{n+1}=0$;
using the first equation at level $n+1$,
$(P_t+E)\psi_{n+1}=2E\,\psi_{n+1}$ --- a recombination valid up
to $O(c^{2})$ corrections, hence exact in the limit --- one obtains
\be
\label{exoeq}
\bigl(P_t-E\bigr)\psi_n=0\,,\qquad
\sqrt{2\kappa_C}\,P_-\psi_n+2E\,\sqrt{n+1}\,\psi_{n+1}=0\,,
\ee
\ie\ the exotic version of the system \eqref{dual}, with the
$n$-dependent coefficient $\sqrt{n+2\alpha}$ replaced by the constant
$\sqrt{2\kappa_C}$: the whole exotic tower sits on the positive-energy
branch $+E$.

Since $(b^-\Psi)_n=\sqrt{n+1}\,\psi_{n+1}$, the
system \eqref{exoeq} takes the compact operator form
\be
\label{exoeqop}
\bigl(P_t-E\bigr)\Psi=0\,,\qquad
\Lambda_-\Psi=0\,,\qquad
\Lambda_-:=\sqrt{2\kappa_C}\,P_-+2E\,b^-\,,
\ee
with $\Lambda_-$ the constraint operator of the exotic Carrollian anyon,
the mirror of its exotic Galilean counterpart \cite{Horvathy:2010vm}.

The solutions of the exotic Carroll equation \eqref{exoeqop} are
readily described.  On a momentum eigenstate,
$P_\pm\Psi=p_\pm\Psi$, the constraint $\Lambda_-\Psi=0$ reads
$b^-\Psi=z\Psi$, with
\be
\label{exosol}
 z(p_-):=-\frac{\sqrt{2\kappa_C}}{2E}\,p_-\,.
\ee
The general solution is therefore
\be\label{ExoSol}
 \Psi=\int d^2p\;\mathcal C(\vec p)\,
 e^{-iEt+i\vec p\cdot\vec x}\,|z(p_-)),
\ee
where
$|z)=e^{-|z|^2/2}\sum_n z^n|n)/\sqrt{n!}$ is the Glauber coherent
state of the internal oscillator, $b^-|z)=z|z)$, and $\mathcal C$ is
an arbitrary momentum-space profile.  The recursion in \eqref{exoeq}
is precisely the coherent-state recursion for the Fock coefficients.
Thus each fixed-momentum mode is a plane wave dressed by an internal
coherent state whose complex label is fixed by $p_-$, while a general
solution is an arbitrary superposition of these
momentum--coherent-state pairs.  For each fixed-momentum mode, the mean
occupation number of the internal oscillator is
\be\label{occupation}
 \bar n(\vec p):=(z|\N|z)
 =\sum_{n=0}^{\infty}n\,|(n|z)|^2
 =|z|^2
 =\frac{\kappa_C|\vec p\,|^2}{2E^2}\,,
\ee
which counts the mean level occupied in the contracted internal spin
tower; it is neither a particle number nor a Landau-level filling
fraction.
This is the same algebraic mechanism that underlies the lowest Landau
level on the ordinary Galilean noncommutative plane
\cite{Horvathy:2002wc,Horvathy:2004fw}.  After projection to the lowest
Landau level, the two guiding-centre coordinates form a Heisenberg pair;
their complex combination is an annihilation operator, and its
eigenstates are Bargmann--Fock coherent states with coefficients
$z^n/\sqrt{n!}$.  Here the contracted spin operators $b^\pm$ provide the
Heisenberg pair, while the polarisation constraint $\Lambda_-\Psi=0$
fixes the coherent-state label to $z(p_-)$ in \eqref{exosol}.  The
analogy is therefore algebraic: no magnetic background or literal
Landau-level projection has yet been introduced.

The internal oscillator is the contracted remnant of the spin tower,
but it no longer carries residual anyonic spin: the latter has been
transmuted into the central charge $\kappa_C$
(\cf~\eqref{exoCasval} below).  The internal rotation spectrum is
integer-spaced,
$\mathrm J=\epsilon_{ij}x_iP_j+\N$, as for the angular-momentum
orbitals within the lowest Landau level.  This makes the algebraic
Bargmann--Fock analogy precise.

\subsection{Exotic Carroll algebra}

Using \eqref{scaledSpin} in the exact boosts \eqref{Kpm} gives
\be
\label{exoboost}
B_\pm=\lim_{c\to0}c^{2}K_\pm
=x_\pm\,P_t\pm i\sqrt{2\kappa_C}\;b^{\pm}\,.
\ee
Both boosts now carry a spin part, and they no longer commute,
\be
\label{exoalg}
[B_+,B_-]=2\kappa_C\,[b^+,b^-]=-2\kappa_C\,,
\qquad\text{\ie}\qquad
[B_i,B_j]=-i\kappa_C\,\epsilon_{ij}\,,
\ee
while the remaining commutators \eqref{Car1}, \eqref{Car2} are unchanged,
with the rotation generator obtained by subtracting the divergent
$c$-number $\alpha=\kappa_C/c^{2}$ from $\mathcal M_0$,
\be
\mathrm J=\epsilon_{ij}x_iP_j+\N\,,\qquad \N=b^{+}b^{-}\,.
\ee
We call the resulting structure --- the Carroll algebra \eqref{Car1},
\eqref{Car2} extended by the central charge \eqref{exoalg} --- the
\emph{exotic Carroll algebra}.  Collecting all the commutation
relations of the generators $\{P_t,P_i,B_i,\mathrm J\}$ in one place,
\bea
&[B_i,P_j]=i\delta_{ij}P_t\,,\qquad
[B_i,B_j]=-i\kappa_C\,\epsilon_{ij}\,,\qquad
[P_i,P_j]=0\,,&
\nonumber\\[4pt]
&[\mathrm J,P_i]=i\epsilon_{ij}P_j\,,\qquad
[\mathrm J,B_i]=i\epsilon_{ij}B_j\,,&
\label{exoticCarrollAlg}\\[4pt]
&[P_t,P_i]=[P_t,B_i]=[\mathrm J,P_t]=0\,,&
\nonumber
\eea
with two central elements: the time-translation operator $P_t$, which
plays the role of the Bargmann mass, and the exotic charge $\kappa_C$,
which measures the non-commutativity of the boosts --- exactly as the
second central charge of the exotic Galilei algebra does for the Galilean
boosts \cite{Lukierski:1996br,Jackiw:2000tz,Duval:2000xr}.

The algebra \eqref{exoticCarrollAlg} is not an ad hoc structure: it fits
precisely into the classification of central extensions of the planar
Carroll group.  While in higher dimensions the Carroll algebra admits no
non-trivial central extensions, in $2+1$ dimensions its second cohomology
has dimension three \cite{deAzcarraga:1997mdw,Matulich:2019cdo}.  The
three cocycles can be supported in the mixed boost--translation, the
boost--boost, and the translation--translation brackets.  The
two-parameter extension used in the coadjoint-orbit construction of
\cite{Marsot:2021tvq} retains the latter two: on top of the already
central time translation, the boosts and spatial translations acquire
central charges,
$[B_i,B_j]\propto\epsilon_{ij}$ (``exotic'') and
$[P_i,P_j]\propto\epsilon_{ij}$ (``magnetic'').  The coadjoint orbits of
this doubly extended group carry four Casimir invariants --- the mass,
the two central charges, and an anyonic spin --- and the classical dynamics of
planar Carroll particles becomes non-trivial in an electromagnetic
background precisely through these charges \cite{Marsot:2021tvq}; the
associated first-order exotic dynamics displays a Peierls substitution and
anomalous Hall motions, and is dual to an uncharged anyon on a black-hole
horizon exhibiting the spin-Hall effect \cite{Zeng:2024bcl}.  In this
language, the contraction \eqref{exoC} provides a field-theoretic,
\.In\"on\"u--Wigner realisation of the exotic (boost-type) extension: the
charge $\kappa_C$ in \eqref{exoalg} is generated by the scaled anyon
spin, with the role of the central mass played by $P_t$, while the
magnetic-type extension remains switched off, $[P_i,P_j]=0$.  This is the
exact anyonic mirror of the way the exotic Galilean charge arises from the
Jackiw--Nair scaling
\cite{Jackiw:2000tz,Duval:2000xr,Horvathy:2004fw}.

The comparison with the coadjoint-orbit analysis of
\cite{Marsot:2021tvq} can be made quantitative.  In the exotic algebra,
the ordinary-Carroll invariant $\widehat{\mathcal C}_2$ of
\eqref{CarCas} is no longer central.  There is no contradiction with
the centrality of the relativistic Pauli--Lubanski scalar:
\eqref{CarCas} was obtained in the fixed-$\alpha$ contraction, for
which the Carroll boosts commute, whereas \eqref{exoC} is a different,
non-uniform contraction.  In the latter, the divergent part
$\alpha=\kappa_C/c^2$ is subtracted from the rotation generator and the
contracted algebra acquires the new cocycle
$[B_i,B_j]=-i\kappa_C\epsilon_{ij}$.  Consequently, the cancellation
that made $P_t\mathrm J-\epsilon_{ij}B_iP_j$ central in the ordinary
Carroll algebra no longer holds; the finite Casimir inherited by the
extended algebra must itself be corrected.  Indeed, \eqref{exoalg}
gives
\be
\bigl[\,P_t\,\mathrm J-\epsilon_{ij}B_iP_j\,,\,B_k\bigr]
=i\kappa_C\,P_k\,,
\ee
and the failure is repaired by a correction quadratic in the momenta,
\be
\label{exoCas}
\CC^{\rm exo}_2:=P_t\,\bigl(P_t\,\mathrm J-\epsilon_{ij}B_iP_j\bigr)
+\frac{\kappa_C}{2}\,\vec P^{\,2}\,,
\qquad
[\CC^{\rm exo}_2,P_k]=[\CC^{\rm exo}_2,B_k]
=[\CC^{\rm exo}_2,\mathrm J]=0\,,
\ee
since $[\frac{\kappa_C}{2}\vec P^{\,2},B_k]=-i\kappa_C\;P_t\,P_k$
compensates the previous commutator once the bracket is multiplied by the
central element $P_t$.  Equation \eqref{exoCas} is the operator
counterpart of the rational Casimir invariant found in
\cite{Marsot:2021tvq} on the coadjoint orbits of the doubly extended
group --- there written, for vanishing magnetic charge, as
$\ell+\vec k\times\vec p/m-q_{2}\,\vec p^{\,2}/m^{2}$ and identified with
the \emph{anyonic spin} of the planar Carroll particle --- multiplied by
the square of the central mass, with $m\to P_t$ and
$q_{2}\propto\kappa_C$.  Its on-shell value follows from
\eqref{exoboost}, which gives
\be
\mathrm J \, P_t-\epsilon_{ij}B_iP_j
=\N\, P_t+\frac{\sqrt{2\kappa_C}}{2}\,\bigl(b^{+}P_-+b^{-}P_+\bigr)\,,
\ee
with the normal-ordered number operator $\N=b^{+}b^{-}$, and from the field
equations \eqref{exoeq}: on solutions $P_t=E$, and the recursion
eliminates $P_-\psi_n$ in favour of $\psi_{n+1}$, whereupon the
contribution of $P_t\,\N$ cancels against that of $b^{+}P_-$, and the
contribution of $b^{-}P_+$ against that of
$\frac{\kappa_C}{2}P_+P_-$.  Hence
\be
\label{exoCasval}
\CC^{\rm exo}_2\,\Psi=0\,.
\ee
The contraction \eqref{exoC} therefore lands on the coadjoint orbits with
$P_t=E$, arbitrary exotic charge $\kappa_C$, and
\emph{vanishing residual anyonic spin}: the whole spin $\alpha$ of the
relativistic anyon has been transmuted into the exotic central charge,
precisely as the Jackiw--Nair scaling trades the anyon spin for the second
central charge of the exotic Galilei algebra
\cite{Jackiw:2000tz,Duval:2000xr,Horvathy:2004fw}.

This trade-off --- the spin sent to infinity while a dimensionful
invariant built from it is held fixed --- is also the mechanism behind the
\emph{continuous-spin} representations of the Poincar\'e group
\cite{Brink:2002zx,Bekaert:2017khg}, which arise from an
\.In\"on\"u--Wigner contraction in which the spin diverges together with a
vanishing mass scale, at fixed dimensionful product
\cite{Khan:2004nj}.  The exotic Carroll (and Galilei) particles may thus
be regarded as planar, non-Lorentzian cousins of the continuous-spin
particle, with the Pauli--Luba\'nski scale replaced by the exotic central
charge $\kappa_C$.

We finalize this section by observing that the exotic limit can also be
carried out with $\alpha\to-\infty$ through the non-unitary
representations, with the same results.

\subsection{Carroll particle in the noncommutative plane}
\label{sec:ncplane}
As in the Galilean case, the exotic central charge has a direct physical
manifestation: the exotic Carroll particle lives on a \emph{noncommutative
plane}.  Let us first recall how this phenomenon arises on the
relativistic and Galilean sides \cite{Horvathy:2006pw}.  In the
relativistic anyon theory the coordinates $x_\mu$ are not observable:
they do not commute with the spin constraint, so acting with them maps
physical states out of the physical subspace.  The observable,
Foldy--Wouthuysen-type \cite{Foldy:1949wa} coordinates are \cite{Cortes:1995wa}
\be\label{bigX}
X_\mu=x_\mu+(P^\rho P_\rho)^{-1}
\epsilon_{\mu\nu\lambda}P^{\nu}\CJ^{\lambda},
\ee
and they are noncommuting on-shell
\cite{Horvathy:2006pw}.  In the Jackiw--Nair limit \cite{Jackiw:2000tz} they become the
coordinates $\mathcal X_i$ of the exotic Galilean particle, which
satisfy
\be
[\mathcal X_i,\mathcal X_j]=i\theta\,\epsilon_{ij}\,,\qquad
\theta=\frac{\kappa_G}{m^{2}}\,,
\ee
so that the theory yields a noncommutative plane
\cite{Horvathy:2002vt,Horvathy:2004fw,Horvathy:2006pw}, the arena of the
Peierls substitution and of the Hall effect
\cite{Duval:2000xr,Duval:2001hu}.

The Carrollian counterpart is immediate, but it is obtained directly
from the already-contracted boost \eqref{exoboost} rather than by
letting $c\to0$ term by term in \eqref{bigX}: the relativistic dressing
\eqref{bigX} divides by $p^{2}$, and under the double scaling
$c\to0,\alpha\to\infty$ its limit is delicate and does not reproduce
the observable coordinate directly, as shown in
Appendix~\ref{app:bigXlimit}.
On the physical subspace \eqref{exoeq}, where $P_t=E$, the boosts
\eqref{exoboost} define the reduced coordinates
\be
\label{XC}
\mathcal X_\pm:=\left.E^{-1}B_\pm\right|_{P_t=E}
=x_\pm\pm i\sqrt{2\theta_C}\,b^{\pm}\,,
\qquad \theta_C:=\frac{\kappa_C}{E^2}\,,
\ee
which are canonically conjugate to the momenta and rotate as a plane
vector, $[\mathcal X_i,P_j]=i\delta_{ij}$,
$[\mathrm J,\mathcal X_\pm]=\pm\mathcal X_\pm$; no momentum-dependent
improvement term is needed, in contrast with the Galilean case
\cite{Horvathy:2006pw}.  These are the \emph{observable} coordinates of
the exotic Carrollian anyon: with the constraint operator $\Lambda_-$ of
\eqref{exoeqop}, one finds
\bea
\label{Xobs}
&[x_+,\Lambda_-]=2i\sqrt{2\kappa_C}\neq0\,,\qquad
[x_-,\Lambda_-]=0\,,&\nonumber\\[4pt]
&[\mathcal X_-,\Lambda_-]=0\,,\qquad
[\mathcal X_+,\Lambda_-]=0\,.&
\eea
Thus $x_+$ takes physical states out of the physical subspace, whereas
$\mathcal X_\pm$ preserve it, exactly like
their relativistic and Galilean forerunners.  Note that
$[\mathcal X_-,\Lambda_-]=0$ holds exactly, since
$[x_-,P_-]=[x_-,b^-]=0$: the coordinate $x_-$ is already an \emph{exact}
observable on its own, in contrast with $x_+$; the whole obstruction to
observability, and hence the whole content of the dressing \eqref{XC},
lies in the $+$ direction.  Their components, however, no longer commute:
\be
\label{XCnc}
[\mathcal X_i,\mathcal X_j]
=-i\theta_{C}\epsilon_{ij}\,.
\ee
The exotic Carrollian anyon lives on a noncommutative plane with
scale $\theta_C$.  This is the Carrollian counterpart of the Galilean
relation $\theta=\kappa_G/m^{2}$, with the Bargmann mass replaced by the
on-shell rest energy $E$; the overall sign tracks that of the exotic
charge in \eqref{exoalg}.
This displays a Landau-like separation between the commuting orbital
variables $x_\pm$ and an internal cyclotron oscillator $b^\pm$.  The
analogy is structural: the internal Fock space is precisely the
Bargmann--Fock module anticipated in remark \textbf{(e)} of
Section~\ref{sec:solutions}, while the noncommutativity belongs to the
dressed observable coordinates $\mathcal X_i$.

The meaning of the coherent-state dressing can now be stated precisely.
For a fixed-momentum mode $\Psi_{\vec p}$, the constraint
$\Lambda_-\Psi_{\vec p}=0$ gives
\[
b^-\Psi_{\vec p}=z(p_-)\Psi_{\vec p}\,,\qquad
z(p_-)=-\sqrt{\frac{\theta_C}{2}}\,p_-\,.
\]
Thus, once $p_-$ is specified, the entire internal spin tower is
determined, up to an overall normalisation, as the coherent state
$|z(p_-))$.  In terms of the noncommutativity scale, the mean occupation
number \eqref{occupation} is
$\bar n(\vec p)=\theta_C|\vec p\,|^2/2$.
Equivalently, the internal part of $\mathcal X_-$ has the definite action
\[
(\mathcal X_--x_-)\Psi_{\vec p}
=-i\sqrt{2\theta_C}\,b^-\Psi_{\vec p}
=i\theta_Cp_-\Psi_{\vec p}\,.
\]
This does not make $\Psi_{\vec p}$ an eigenstate of the complete
coordinate $\mathcal X_-$: the orbital operator $x_-$ remains
non-diagonal on a momentum eigenstate.  Nor is a coherent state an
eigenstate of $b^+$.  Accordingly, and consistently with
$[\mathcal X_1,\mathcal X_2]=-i\theta_C$, the two observable
coordinates admit no joint eigenstates.  One difference with the Galilean case
deserves emphasis.  There, the coordinates $x_i$ undergo a
Zitterbewegung-like motion, and the $\mathcal X_i$ are singled out by
their free evolution $\dot{\mathcal X}_i=p_i/m$
\cite{Horvathy:2004fw,Horvathy:2006pw}.  In the Carrollian case the
first equation in \eqref{exoeqop} determines the time evolution in terms of the constant Hamiltonian function $E$.  Thus it contains no momentum-dependent
kinetic term, and the Heisenberg equations give
\be
\label{staticity}
\dot x_i=i[E,x_i]=0\,,\qquad
\dot{\mathcal X}_i=i[E,\mathcal X_i]=0\,.
\ee
 This is consistent with the Carroll algebra: the central generator
$P_t$ commutes with $x_i$ and with the internal oscillators entering
\eqref{XC}.  The staticity of the Carroll particle therefore holds
equally in the original and in the Foldy--Wouthuysen-type coordinates.

Equivalently, the general wave function \eqref{ExoSol} has the stationary
form $\Psi(t,\vec x)=e^{-iEt}\Psi(0,\vec x)$: both the momentum profile
$\mathcal C(\vec p)$ and the internal coherent-state dressing
$|z(p_-))$ are time-independent.  Hence every momentum component carries
the same overall phase, which cancels from expectation values of
observables with no explicit time dependence.  In particular, free
Carroll particles do not move \cite{Marsot:2021tvq}.  The coordinates
\eqref{XC} are therefore singled out on purely kinematical grounds: they
are simultaneously observable and noncommutative.

\subsection{The Carrollian time coordinate}
\label{sec:carrolltime}

Finally, let us consider the \emph{time} component of the covariant
coordinates.  In the exotic Carroll limit both terms of $X^{0}$ are of
order $c$: the orbital part $x^{0}=ct$ trivially, and the spin part
because $c\CJ_\pm\to\sqrt{2\kappa_C}\,b^{\pm}$ by
\eqref{scaledSpin}, while
$(P^\rho P_\rho)^{-1}\to-c^{2}P_t^{-2}$.  The same renormalisation that turns $x^{0}$
into $t$ therefore leaves a finite and nontrivial kinematical time
coordinate.  Indeed,
\eqref{bigX} gives
$X^0=x^0+\frac{i}{2}(P^\rho P_\rho)^{-1}
(P_+\CJ_- -P_-\CJ_+)$, and hence
\be
\label{Tdef}
\mathcal T:=\lim_{c\to0}\frac{X^{0}}{c}
=t-\frac{i\sqrt{2\kappa_C}}{2}\,P_t^{-2}
\bigl(P_+b^{-}-P_-b^{+}\bigr)\,.
\ee
$\mathcal T$ is Hermitian, a scalar under rotations, commutes with the
momenta, and is canonically conjugate to the central energy,
$[\mathcal T,P_t]=-i$.  In the Galilean mirror
the analogous correction is suppressed by $1/c^{2}$ and the limit gives
$\mathcal T_{G}=t$ exactly: \emph{absolute Galilean time survives its
limit undeformed, whereas the covariant Carrollian time is
operator-valued} --- as befits the two degenerate structures, in which
Galilei keeps an absolute time and Carroll an absolute space, so that it
is the Carrollian time fibre that becomes fuzzy.
Because $\mathcal T$ changes the energy, its mixed commutators must be
computed before restricting to $P_t=E$.  For this purpose define the
off-shell lift of \eqref{XC},
\be
\mathscr X_\pm:=P_t^{-1}B_\pm\,,\qquad
\left.\mathscr X_\pm\right|_{P_t=E}=\mathcal X_\pm\,.
\label{Xoff}
\ee
A direct operator calculation gives
\be
\label{TXnc}
[\mathcal T,\mathscr X_i]=
\frac{i\,\kappa_C}{P_t^{3}}\,\epsilon_{ij}P_j
\;\approx\;\frac{i\theta_C}{E}\,\epsilon_{ij}P_j\,.
\ee
The direct exotic Carroll limit of the relativistic relation
$[X_\mu,X_\nu]\approx-i\,s\,\epsilon_{\mu\nu\lambda}
p^{\lambda}(-p^{2})^{-3/2}$ of \cite{Cortes:1995wa,Horvathy:2006pw}
is instead carried by the Bopp-shifted coordinate~\cite{Bopp:1956}
$X_i^{\mathrm{lim}}=\mathscr X_i-\kappa_C P_t^{-2}
\epsilon_{ij}P_j$ derived in Appendix~\ref{app:bigXlimit}; its
time--space bracket has the opposite sign.  Equation~\eqref{TXnc} is the
bracket of the off-shell lift \eqref{Xoff} of the observable coordinates
\eqref{XC}.  It is compatible with the Jacobi identity because
$[\mathscr X_i,\mathscr X_j]=-i\kappa_C P_t^{-2}\epsilon_{ij}$ and
$[\mathcal T,P_t]=-i$; its restriction to $P_t=E$ reproduces
\eqref{XCnc}.  The resulting noncommutative geometry is thus
two-tiered: a constant spatial cell \eqref{XCnc}, and a
momentum-dependent time--space cell \eqref{TXnc}.
One subtlety distinguishes time from space.  Unlike $\mathcal X_\pm$,
the operator $\mathcal T$ does not preserve the kernel of the internal
constraint, $[\mathcal T,\Lambda_-]
=-i\sqrt{2\kappa_C}\,E\,P_t^{-2}P_-$,
whereas its minimal $\N$-improvement does so \emph{exactly},
\be
\label{TLam}
[\mathcal T',\Lambda_-]=-iE\,P_t^{-2}\Lambda_-\,,
\qquad
\mathcal T':=\mathcal T+iE\,P_t^{-2}\N\,,
\ee
although $\mathcal T'$ is not Hermitian.  The significance of
$\mathcal T'$ is not that it defines a physical time observable of the
fixed-energy system: conjugacy to $P_t$ gives
$[\mathcal T',P_t-E]=-i$ and
$[\mathcal T',\mathcal D_C]=-2iP_t$, so it preserves neither the energy
shell nor the Carroll equation.  Rather, \eqref{TLam} shows that
$\mathcal T'$ is a constraint-adapted lift of the time coordinate,
motivated directly by the complex polarisation selected by
$\Lambda_-\Psi=0$.  The imaginary number-operator correction is thus a
nontrivial quantisation rule on the polarised space.

\subsection{The gapless boundary of the exotic family}
\label{sec:exogapless}

The exotic scaling may also be performed at fixed \emph{mass}, so that
$E=mc^{2}\to0$.  The divergence of the coherent-state label
$z=-\sqrt{2\kappa_C}\,p_-/2E$ signals that this limit is delicate, and
indeed it bifurcates.  Keeping $\kappa_C$ fixed, the limit of
\eqref{exoexact} gives $P_t\psi_n=0$ \emph{and}
$P_-\psi_n=0$: a fully static tower supported on the kernel of $P_-$,
with no recursion linking the levels and the internal Fock space left
unconstrained.  Equivalently, the imposed constraint degenerates to
$\Lambda_-\to\sqrt{2\kappa_C}\,P_-$, while its adjoint tends to
$\Lambda_+\to\sqrt{2\kappa_C}\,P_+$. Thus the vanishing commutator
$[\Lambda_-,\Lambda_+]=4E^{2}\to0$ records the degeneration of the
second-class polarisation.  At the same time, the noncommutativity parameter
$\theta_C=\kappa_C/E^{2}$ of \eqref{XCnc} diverges: an infinitely
noncommutative, lowest-Landau-level-like collapse onto the chiral sector
of remark \textbf{(e)} of Section~\ref{sec:solutions}.

There is also a second, singular boundary within the
already-contracted family.  Keeping the noncommutativity scale
$\theta_C$ of \eqref{XCnc} finite as $E\to0$ requires
$\kappa_C=\theta_C E^2\to0$.  After dividing the constraint by $2E$,
one obtains the static coherent system
\be
\label{thetaSys}
P_t\psi_n=0\,,\qquad
\sqrt{\frac{\theta_C}{2}}\,P_-\psi_n
+\sqrt{n+1}\,\psi_{n+1}=0\,,
\ee
Equivalently,
\be
\label{thetaSysOp}
P_t\Psi=0\,,\qquad
\Gamma_-\Psi=0\,,\qquad
\Gamma_-:=\sqrt{\frac{\theta_C}{2}}\,P_-+b^-\,.
\ee
The momentum eigenstates of \eqref{thetaSysOp} remain dressed by
coherent states of finite
label $z=-\sqrt{\theta_C/2}\,p_-$.  The exotic
cocycle vanishes and the boosts reduce to the ordinary commuting Carroll
ones.  Both ordinary Carroll Casimirs \eqref{CarCas} vanish on the
physical module, $
\widehat{\mathcal C}_1\Psi=0$, $
\widehat{\mathcal C}_2\Psi=0$. 
Indeed, $B_i=x_iP_t$ and
$\mathrm J=\epsilon_{ij}x_iP_j+\N$ give
$\widehat{\mathcal C}_2=\N P_t$, which vanishes on \eqref{thetaSysOp}.
The oscillator still enters $\mathrm J$ and the component recursion, but
it does not furnish a residual Carroll-spin eigenvalue.  This differs
from the massless planar anyons of Ref.~\cite{Marsot:2021tvq}, which
retain non-zero extension charges and an independent anyonic-spin
Casimir; here both $P_t$ and $\kappa_C$ vanish.  The comparison with the
coadjoint-orbit spin therefore applies to the finite-energy exotic
family, not directly to this boundary.

Although $\theta_C$ remains finite by construction, it is no longer
fixed by a non-zero central charge of the limiting algebra.  The
coordinates \eqref{XC} have the finite trajectory limit
\[
\mathcal X^{(0)}_\pm
:=\lim_{\substack{E\to0\\ \theta_C\ {\rm fixed}}}\mathcal X_\pm
=x_\pm\pm i\sqrt{2\theta_C}\,b^\pm\,.
\]
Their Bopp-shifted counterparts
\[
q_\pm=\mathcal X^{(0)}_\pm\pm\frac{i\theta_C}{2}P_\pm
\]
commute with one another and with the reduced constraint operator,
$[q_\pm,\Gamma_-]=0$, and hence preserve its kernel.  More explicitly,
setting
\[
a:=\sqrt{\frac{\theta_C}{2}}\,,\qquad
U_a:=\exp\!\left[a\bigl(P_-b^+-P_+b^-\bigr)\right]
\]
gives the exact relations
\[
\Gamma_-=U_a^{-1}b^-U_a\,,\qquad
q_\pm=U_a^{-1}x_\pm U_a\,.
\]
Thus, for $\widetilde\Psi=U_a\Psi$, the coherent equation becomes
$b^-\widetilde\Psi=0$.  The fixed-$\theta_C$ system is a
momentum-dependent coherent realisation of the zero-spin gapless
Carroll module, rather than a new spin sector.

The fixed-$\kappa_C$ and fixed-$\theta_C$ procedures therefore define two
distinct gapless boundaries of the finite-energy family.  The former
retains the exotic Carroll algebra but yields the chiral condition
$P_-\Psi=0$, with the internal tower left unconstrained; the latter
retains the coherent momentum--oscillator recursion but has
$\kappa_C\to0$. 

%%%%%%%%%%%%%%%%%%%%%%%%%%%%%%%%%%%%%%%%%%%%%%%%%%%%%%%%%%%%%%%%%%%%%%%%%%%%%%
\section{Galilean--Carroll branching from a common similarity
transformation}
\label{sec:branching}
%%%%%%%%%%%%%%%%%%%%%%%%%%%%%%%%%%%%%%%%%%%%%%%%%%%%%%%%%%%%%%%%%%%%%%%%%%%%%%

The Carroll limit of the main text required only the grading \eqref{SC}.
Historically, however, the non-Lorentzian limit of the anyon system was
first taken on the Galilean side \cite{Horvathy:2010vm}, with a
similarity transformation that combines the same grading with a
time-dependent rest-energy phase.  In this section we show how
\emph{both} non-Lorentzian descendants --- the Galilean anyon and the
Carrollian anyon of Section~\ref{sec:carroll} --- branch from that
single phase-dressed transformation, and we draw the corresponding
lesson on the fate of the two relativistic energy branches.

For the free scalar the role of the phase is familiar.  In the Galilean
regime $c\to\infty$ the rest-energy term $m^{2}c^{4}$ of
\eqref{KGscalar} diverges: the solutions oscillate as
$e^{\mp imc^{2}t}$ and no naive limit exists.  The standard cure
\cite{Aldaya:1985gy,Duval:2002cw} is a time-dependent phase
redefinition, $\phi=e^{+imc^{2}t}\psi$, after which \eqref{KGscalar}
becomes
$\bigl(\partial_t^{2}-2imc^{2}\partial_t-c^{2}\vec\nabla^{2}\bigr)\phi=0$;
dividing by $2mc^{2}$ and letting $c\to\infty$ at fixed $m$ yields the
free Schr\"odinger equation,
$\bigl(P_t+\tfrac{1}{2m}\vec\nabla^{2}\bigr)\phi=0$.  The phase
strips the rest-energy oscillation of the \emph{positive}-energy branch:
it is adapted to the branch that the Galilean limit keeps.  By contrast,
in the phase-free Carroll contraction of Section~\ref{sec:SC}, both
relativistic rest-energy branches survive at finite Carroll frequencies,
$P_t=\pm E$.

\subsection{The common transformation and the Galilean anyon}
\label{sec:galilean}

For the anyon system, the transformation of \cite{Horvathy:2010vm} is
\be
\label{SG}
\Phi_G:=S_G\,\psi\,,\qquad
S_G:=e^{+imc^{2}t}\,c^{\,\N}\,,\qquad
\N:=\CJ_0-\alpha\,,
\ee
\ie\ the grading \eqref{SC} dressed by the Galilean rest-energy phase.
Following the regularised contraction of
\cite{Aldaya:1985gy,Duval:2002cw}, the phase should be viewed as
adjoining a trivial central factor --- the mass operator --- to the
Poincar\'e algebra before taking the limit; the algebra being
contracted is the trivial central extension
$\widehat{\mathfrak{iso}}(2,1)=\mathfrak{iso}(2,1)\oplus\mathfrak
u(1)_m$.  On the time-translation generator the phase acts as
$S_G\,P^{0}S_G^{-1}=P_t/c+mc$, so that
$mc-S_GP^{0}S_G^{-1}=-\frac1c\,P_t$ and
$mc+S_GP^{0}S_G^{-1}=\frac1c\,(P_t+\Epsilon)$, with
\be
\Epsilon:=2mc^{2}=2E
\ee
the interbranch gap.  The exact component equations \eqref{HP1},
\eqref{HP2} become
\be
\label{GframeConjugate}
\sqrt{n+2\alpha}\;P_t\,\phi_n+\sqrt{n+1}\,P_+\phi_{n+1}=0\,,
\qquad
c^{2}\sqrt{n+2\alpha}\,P_-\phi_n
+\sqrt{n+1}\bigl(P_t+\Epsilon\bigr)\phi_{n+1}=0\,.
\ee
The two frames are related by $\Phi_G=e^{+iEt}\,\Phi$: comparing with
\eqref{I-exact}, \eqref{II-exact}, the phase has shifted every frequency
by $-E$, turning the symmetric pair $P_t\mp E$ into
$P_t$ and $P_t+\Epsilon$.

Dividing the second equation of \eqref{GframeConjugate} by $c^{2}$ and
letting $c\to\infty$ at
fixed $m$ --- so that $\Epsilon/c^{2}=2m$, while the time derivative is
suppressed by $1/c^{2}$ --- reproduces the Galilean anyon
system of \cite{Horvathy:2010vm},
\be
\label{GalileicompConjugate}
\sqrt{n+2\alpha}\;P_t\,\phi_n+\sqrt{n+1}\,P_+\phi_{n+1}=0\,,\qquad
\sqrt{n+2\alpha}\,P_-\phi_n+2m\sqrt{n+1}\,\phi_{n+1}=0\,.
\ee
This is the arbitrary-spin generalisation of the L\'evy--Leblond equations
\cite{Levy-Leblond:1967eic}.  The second equation of
\eqref{GalileicompConjugate} contains no time derivative: it is a purely
algebraic (auxiliary) relation,
\be
\label{GalrecurConjugate}
\phi_{n+1}=-\frac{1}{2m}\sqrt{\frac{n+2\alpha}{n+1}}\;P_-\phi_n\,,
\ee
so the whole spin tower is generated \emph{upwards} (increasing tower level
$n$) from the single dynamical component $\phi_0$ by repeated application
of $P_-$.  Substituting \eqref{GalrecurConjugate} back into the first
equation of \eqref{GalileicompConjugate} eliminates $\phi_{n+1}$, and
since $P_+P_-=\vec P^{\,2}$, \emph{every} component of the tower obeys
the free Schr\"odinger equation,
\be
\label{SchrodCompConjugate}
\Bigl(P_t-\frac{\vec P^{\,2}}{2m}\Bigr)\phi_n=0\,,
\qquad n=0,1,2,\dots
\ee
The system \eqref{GalileicompConjugate} is thus a first-order ``square
root'' of the Schr\"odinger equation, and it carries an irreducible
spin-$\alpha$ representation of the Bargmann (centrally extended
Galilei) algebra, with Casimir eigenvalues $\CC_1=0$ and
$\CC_2=m\alpha$ \cite{Horvathy:2010vm}.

For a momentum eigenstate, iteration of \eqref{GalrecurConjugate} gives
\be
\label{GalTowerSolution}
\phi_n=
\left(-\frac{p_-}{2m}\right)^n
\sqrt{\frac{(2\alpha)_n}{n!}}\;\phi_0\,.
\ee
For $\alpha>0$, the internal-module norm at fixed momentum is
\be
 \sum_{n=0}^{\infty}|\phi_n|^2
 =|\phi_0|^2\sum_{n=0}^{\infty}
 \frac{(2\alpha)_n}{n!}
 \left(\frac{|\vec p\,|^2}{4m^2}\right)^n
 =|\phi_0|^2
 \left(1-\frac{|\vec p\,|^2}{4m^2}\right)^{-2\alpha},
\ee
By the ratio test, this series converges precisely for
$|\vec p\,|<2m$, where the last equality holds.   In
particular, this condition includes the rest frame: at $\vec p=0$,
\eqref{GalrecurConjugate} sets all $\phi_{n\geq1}$ to zero and leaves
the lowest component with vanishing Schr\"odinger energy.  The
Carrollian tower instead requires $|\vec p\,|>2E$, although its
positive-energy lowest component exists at every momentum, as explained
in remarks~\textbf{(b)} and \textbf{(d)} of
Section~\ref{sec:solutions}.

The corresponding Galilean boosts are
\be
\CK_\pm=-tP_\pm+mx_\pm+\Delta_\pm\,,
\qquad
\Delta_+=i\CJ_+\,,\qquad \Delta_-=0\,.
\ee
In the Carrollian mirror \eqref{Bpm}, by contrast, the internal spin
block is carried by the opposite boost: $B_+=x_+P_t$ and
$B_-=x_-P_t-i\CJ_-$.  Thus the spin contribution migrates from
$\CK_+$ in the Galilean contraction to $B_-$ in the Carrollian
contraction.

\subsection{The Carroll limit in the Galilean frame, and the lesson}

Nothing forbids taking the \emph{Carroll} limit in the same \eqref{SG} field basis.
Letting $c\to0$ at fixed $\Epsilon=2mc^{2}$ in \eqref{GframeConjugate},
the first equation survives untouched while the second reduces to
$(P_t+\Epsilon)\phi_n=0$ for $n\geq1$; inserting this branch
condition into the first equation returns the recursion
\be
\label{recurConjugate}
P_+\phi_{n+1}=\Epsilon\,\sqrt{\tfrac{n+2\alpha}{n+1}}\;\phi_n\,,
\qquad n\geq1\,,
\ee
the Carrollian analogue of the Galilean auxiliary relation
\eqref{GalrecurConjugate}, and the full tower obeys
\be
\label{CarrollConjugate}
P_t\bigl(P_t+\Epsilon\bigr)\phi_n=0\,,
\ee
with the two energy branches $\{-\Epsilon,0\}$: the relativistic positive-energy
mode is \emph{frozen} at zero frequency, while the negative-energy,
past-directed mode keeps propagating at $-\Epsilon$.  This is the
system of Section~\ref{sec:carroll} in disguise: multiplying by
$e^{-iEt}$ recentres the spectrum at $\pm E$.

Had we used the conjugate phase $e^{-imc^{2}t}$ in \eqref{SG} instead, the Carrollian
energies would come out non-negative, $\{0,+\Epsilon\}$ --- the
convention of \cite{Parekh:2026wri} --- but the \emph{Galilean} limit
taken in that frame produces the Schr\"odinger equation with a curious
additive constant,
\be
P_t\,\phi_n=\Bigl(\Epsilon+\frac{\vec P^{\,2}}{2m}\Bigr)\phi_n\,.
\ee
The Galilean interpretation is now clear: that phase subtracts the rest energy
of the \emph{negative}-energy branch, so the zero of energy is set at
the bottom of the relativistic gap, and the surviving Galilean particle
consequently lies one full interbranch gap $\Epsilon=2mc^{2}$ above the
chosen zero of energy, in addition to its kinetic energy.  Each phase,
$\exp(\pm imc^2t)$, is adapted to one relativistic branch.  For the
Carroll contraction, by contrast, the phase-free frame used in the main
text places the two surviving branches symmetrically.

At the formal level all these frames are strictly equivalent.  In the
frame shifted by $U=e^{-iEt}=e^{-imc^{2}t}$ relative to the main text,
the Carroll equation \eqref{CarrollCasEq} reads
\be
\label{CarrollShifted}
\partial_t\bigl(\partial_t+i\Epsilon\bigr)\Phi'=0\,,\qquad
\Phi'=U\,\Phi\,,
\ee
with the non-negative spectrum $\{0,\Epsilon\}$, and the two quadratic
equations are one and the same equation written in terms of the
transformed time-translation operator \cite{Parekh:2026wri}
\be
\label{Dtdef}
\Dt:=U\,\partial_t\,U^{-1}=\partial_t+\frac{i\Epsilon}{2}\,,
\qquad
(i\Dt)^{2}\Phi'=E^{2}\,\Phi'\,.
\ee
Since $U$ is an invertible, purely time-dependent field redefinition,
the two systems of equations are strictly equivalent: the frames differ
only in the choice of field basis.  More precisely, since
$e^{\tau\Dt}=e^{i\Epsilon\tau/2}\,e^{\tau\partial_t}$, the two
generators $\Dt$ and $\partial_t$ are two \emph{lifts} of one and the
same projective time translation, differing by a central $U(1)$ phase
--- the $\mathfrak u(1)_m$ factor above --- which is unobservable on
individual states.  What is observable is the \emph{relative} phase
accumulated between the two energy branches in a lapse $\tau$,
$e^{-i\Epsilon\tau}$, and it is the same in every frame: the invariant
content of Carrollian time evolution is the gap, not the choice of
energy origin.

The lesson of the common transformation \eqref{SG} is that the two
non-Lorentzian limits are complementary fates of the two relativistic
energy branches.  The \emph{Galilean} limit gets rid of the
negative-energy, past-directed modes --- they recede to infinite
negative energy and are traded for the auxiliary relation --- keeping a
single future-directed mode governed by the Schr\"odinger equation.
The \emph{Carroll} limit, by contrast, discards neither relativistic
energy branch: in the phase-dressed frame it freezes the branch selected
by the phase and retains the other at finite frequency, so both rest
energies survive the contraction.  What is lost is the momentum
dependence of the relativistic dispersion relation.  The spatial
momentum remains a label of the wave function and of the spin-tower
recursion, but it no longer contributes to the energy, and all momenta
within either branch become degenerate.  The two-sector structure of
the Carrollian anyon --- the positive-energy mode $\chi$ and the
negative-energy tower of Section~\ref{sec:solutions} --- is therefore
the Carrollian remnant of the relativistic particle--antiparticle
doubling, of which the Galilean theory keeps only half.

%%%%%%%%%%%%%%%%%%%%%%%%%%%%%%%%%%%%%%%%%%%%%%%%%%%%%%%%%%%%%%%%%%%%%%%%%%%%%%
\section{Conclusions and outlook}
\label{sec:conclusions}
%%%%%%%%%%%%%%%%%%%%%%%%%%%%%%%%%%%%%%%%%%%%%%%%%%%%%%%%%%%%%%%%%%%%%%%%%%%%%%

The relativistic anyon admits a remarkably clean Carrollian
contraction: a pure grading of its spin-tower components by powers of
$c$, the similarity transformation \eqref{SC}, taken at fixed rest
energy $E=mc^{2}$.  The resulting vector constraints close on the
central secondary operator $\mathcal D_C=P_t^2-E^2$ and therefore form
an arbitrary-spin, first-order square root of the Carroll equation.  At
$\alpha=-j$ the system truncates to finite-component Carrollian fields:
spin-$\tfrac12$ gives the Carrollian L\'evy--Leblond equations, spin-$1$
the Carroll limit of the Deser--Jackiw--Templeton system, and spin-$\tfrac32$ a finite-energy Carrollian Rarita--Schwinger system, with the
pattern extending  to arbitrary (half-)integer spin in a precise way.

This contraction extends to the $c\to0$ regime the strategy used in
\cite{Horvathy:2010vm} for the Galilean limit of the same system, and
the comparison of the two regimes is instructive in itself
(Section~\ref{sec:branching}): both
non-Lorentzian theories branch from one common transformation, but they
treat the two relativistic rest-energy branches in opposite ways.  The
Galilean limit expels the negative-energy, past-directed modes ---
they recede to infinitely negative energy and are traded for the
auxiliary relation \eqref{GalrecurConjugate} --- whereas the Carroll
limit discards neither branch.  What it does erase is their momentum
dependence: spatial momentum continues to label the wave functions and
the spin-tower recursion, but no longer contributes to the energy.
What remains of the relativistic dispersion relation when the light
cones collapse is precisely its two rest energies.  Every component
obeys the Carroll equation
$(\partial_t^{2}+E^{2})\phi=0$, with two flat branches $\pm E$
separated by the gap $\Delta E=2E$ that no
phase redefinition can remove.  The two-sector structure of the
Carrollian theory is thus the $c\to0$ remnant of the relativistic
particle--antiparticle doubling, of which the Galilean theory keeps
only half.

The complementarity also appears in the convergence of the internal
spin towers.  The Galilean solution is an expansion in
$|\vec p\,|/(2m)$ and converges for $|\vec p\,|<2m$, whereas the
Carrollian tower is controlled by the reciprocal-type ratio
$2E/|\vec p\,|$ and converges for $|\vec p\,|>2E$.  For the unitary
discrete series $\alpha>0$ the corresponding critical boundaries are
excluded.  At the Carrollian boundary $|\vec p\,|=2E$, however, the
ratio test becomes inconclusive and the power-law tail remains summable
in the first non-unitary interpolation window
$-\tfrac12<\alpha<0$.

The resulting particle combines a flat Carrollian dispersion with a
non-trivial internal structure.  Within either frequency branch, every
momentum component acquires the same time-dependent phase, so a wave
packet has no momentum-dependent free propagation; a superposition of
the two branches nevertheless accumulate a non-trivial relative
phase.  Meanwhile, its spin --- the arbitrary real Carroll invariant
$\widehat{\mathcal C}_2=E\alpha$ --- is carried by a tower of components
tied together by the recursion \eqref{recur}, so that
each branch
propagates a single independent mode.

When the spin is scaled along with $c$, $\kappa_C=\alpha c^{2}$ fixed,
in the Carrollian analogue of
the Jackiw--Nair limit, the fractional spin is not lost but
\emph{transmuted}: it reappears as the second central charge of
$[B_i,B_j]=-i\kappa_C\,\epsilon_{ij}$, and thereby as a physical
noncommutative area scale --- the observable coordinates of the exotic Carroll
particle close on a
noncommutative plane with scale $\theta_C=\kappa_C/E^{2}$, mirroring the
exotic Galilean relation $\theta=\kappa_G/m^{2}$ of
\cite{Duval:2000xr,Horvathy:2002vt,Horvathy:2006pw} with the mass
replaced by the rest energy.  The states themselves organise into
lowest-Landau-level-like coherent states \eqref{ExoSol}.  Moreover, the
covariant relativistic time coordinate admits the finite,
operator-valued Carroll limit \eqref{Tdef}, although it is not a physical
time observable within a fixed-energy sector.  The energy gap is what
holds this structure
together: at vanishing gap one may retain the exotic algebra or the
coherent dynamics, but not both.

Several directions suggest themselves.  The most inviting one is the
coupling to magnetic backgrounds.  The holomorphic sector of remark
\textbf{(e)} is lowest-Landau-level-like: its zero modes converge along
the internal spin-tower direction but are not square-integrable over the
free spatial plane.  A magnetic background can supply the Gaussian
Bargmann--Fock measure that renders such holomorphic modes normalisable
--- and magnetic backgrounds are also what sets
planar Carroll particles in motion
\cite{Marsot:2021tvq,Zeng:2024bcl}, with the observable coordinates
\eqref{XC} as the natural variables for a Peierls-type coupling.
It is therefore natural to ask whether a contraction in such a
background can also switch on the magnetic central extension, completing
the doubly extended Carroll algebra of \cite{Marsot:2021tvq}.  Magnetic
translations are natural candidates for supplying the second central
charge, and one may expect the magnetically coupled wave equations to
provide a dynamical realisation of the resulting algebra.  Establishing
this construction is left for future work.  If realised, it could also
sharpen, but would not by itself settle, the fate of fractional
statistics.  The
exchange phase of identical particles is a topological rather than
metric datum, and the exotic contraction fixes $\kappa_C=\alpha c^2$
but not the fractional part of the parent spin.  The braid phase must
therefore be retained as additional global data, while ultra-locality
trivialises its local dynamics.  A Carrollian Chern--Simons coupling
seems the natural way to make this precise, also in view of the
quantum-information applications of anyon braiding
\cite{Kitaev:1997wr,Nayak:2008zza,Andersen:2022xmz}.

On the structural side, it would be natural to construct actions for
the Carrollian anyon --- identifying its ``electric'' and ``magnetic''
sectors along the lines of the fermionic actions of
\cite{Parekh:2026wri} --- and to classify boost-compatible
interactions, which ultra-locality should render far more tractable
than in the Galilean theory.  The supersymmetric aspects should extend
naturally as well: the Carrollian systems constructed here are expected
to possess supersymmetric extensions, in parallel with their Galilean
mirrors \cite{Horvathy:2010vm}.  Constructing and classifying the
corresponding super-Carroll wave systems is left for future work.

Finally, the exotic particle hides an intriguing structure in its time
direction: on fixed-momentum states the enlarged Fock module organises
into a formal ladder in complexified time, with imaginary spacing set
by the inverse energy gap, of which the physical constraint selects the lowest
level.  Clarifying whether this complex time structure acquires
a genuine physical role, when transitions between energy sectors or
interactions are allowed, is an intriguing open problem.

%%%%%%%%%%%%%%%%%%%%%%%%%%%%%%%%%%%%%%%%%%%%%%%%%%%%%%%%%%%%%%%%%%%%%%%%%%%%%%
\section*{Acknowledgements}
%%%%%%%%%%%%%%%%%%%%%%%%%%%%%%%%%%%%%%%%%%%%%%%%%%%%%%%%%%%%%%%%%%%%%%%%%%%%%%

This work is supported by ANID Fondecyt grants N$^{\circ}$1252053,
N$^{\circ}$1250672, N$^{\circ}$1220862.

%%%%%%%%%%%%%%%%%%%%%%%%%%%%%%%%%%%%%%%%%%%%%%%%%%%%%%%%%%%%%%%%%%%%%%%%%%%%%%
\appendix
%%%%%%%%%%%%%%%%%%%%%%%%%%%%%%%%%%%%%%%%%%%%%%%%%%%%%%%%%%%%%%%%%%%%%%%%%%%%%%

\section{Conventions and algebraic identities}
\label{app:conv}
The spacetime conventions are those of Section~\ref{sec:setup}:
$\eta_{\mu\nu}=\mathrm{diag}(-1,1,1)$, $x^{0}=ct$,
$P_\mu=-i\partial_\mu$, and $P^{0}=P_t/c=-P_0$.  With
$\mathcal M_\pm=\mathcal M_1\pm i\mathcal M_2$ and
$P_\pm=P_1\pm iP_2$, the nonvanishing commutation relations of the
Poincar\'e algebra \eqref{PoincareAlg} are
\bea
&[\mathcal M_0,\mathcal M_\pm]=\pm\mathcal M_\pm\,,\qquad
[\mathcal M_+,\mathcal M_-]=-2\mathcal M_0\,,&
\nonumber\\[3pt]
&[\mathcal M_0,P_\pm]=\pm P_\pm\,,\qquad
[\mathcal M_\pm,P^{0}]=\pm P_\pm\,,\qquad
[\mathcal M_\pm,P_\mp]=\pm2P^{0}\,.&
\label{PoincarePM}
\eea
The corresponding coordinate combinations $x_\pm=x_1\pm ix_2$ obey
$[x_\pm,P_\mp]=2i$ and $[x_\pm,P_\pm]=0$.  With
$\epsilon_{12}=1$, two planar vectors satisfy
$\epsilon_{ij}U_iV_j=\frac{i}{2}(U_+V_--U_-V_+)$ and
$(\epsilon_{ij}U_j)_\pm=\mp iU_\pm$; consequently
$[\mathrm J,X_\pm]=\pm X_\pm$ for any planar vector $X$.  The grading
\eqref{SC} leaves $P_t$, $P_i$, and $E=mc^2$ unchanged.

The spin generators $\CJ_\pm$ and $\CJ_0$ satisfy the same Lorentz algebra.  Its half-bounded representations are given by
\eqref{Dplus}.  The descendant notation used there follows by iterating
the raising-operator action:
\be
(\CJ_+)^n|0)
=\prod_{k=0}^{n-1}C_k^\alpha\,|n)
=\sqrt{n!\,(2\alpha)_n}\,|n)\,,
\qquad
(a)_n:=a(a+1)\cdots(a+n-1)\,,\qquad (a)_0:=1\,.
\label{VermaDescendants}
\ee
Thus, for $\alpha>0$ and $(0|0)=1$, the normalised states are
\be
|n)=\frac{1}{\sqrt{n!\,(2\alpha)_n}}(\CJ_+)^n|0)\,.
\label{VermaNormalisation}
\ee
For negative $\alpha$ the proportionality remains algebraically valid,
but the invariant form is indefinite and the square roots require the
phase conventions described below.

The finite-dimensional truncation can be seen before taking the quotient
of the lowest-weight Verma module.  If
$v_n:=(\CJ_+)^n|0)$, the Lorentz algebra gives
\be
\CJ_-v_n=n(2\alpha+n-1)v_{n-1}\,,\qquad
\CJ_0v_n=(\alpha+n)v_n\,.
\label{VermaLowering}
\ee
At $\alpha=-j$, the level-$2j+1$ state satisfies
$\CJ_-v_{2j+1}=0$ and has zero invariant norm, since
$(2\alpha)_{2j+1}=(-2j)_{2j+1}=0$.  It is therefore a singular
lowest-weight vector, and it generates, together with all its descendants,
a null invariant submodule.  Quotienting the Verma module by this
submodule leaves the states $|0),\ldots,|2j)$; the image of $|2j)$ is a
highest-weight state because $\CJ_+|2j)=0$ in the quotient.

For $\alpha=-j$ the
coefficients $C^{-j}_n$ are imaginary for $n<2j$; redefining
$\CJ_\pm\to\pm i\CJ_\pm$ one gets
\be
\label{C'D}
\CJ_+|n)=C'^{-j}_n|n+1)\,,\qquad
\CJ_-|n)=-C'^{-j}_{n-1}|n-1)\,,\qquad
C'^{-j}_n=\sqrt{(2j-n)(n+1)}\,,
\ee
with the indefinite metric $\eta_{nn'}=\mathrm{diag}(1,-1,1,\dots)$.  In
the component (column) representation,
$(\CJ_+\Phi)_n=C^\alpha_{n-1}\phi_{n-1}$ and
$(\CJ_-\Phi)_n=C^\alpha_n\phi_{n+1}$, so that $c^{\,\N}$ gives
$\widetilde\CJ_\pm=c^{\pm1}\CJ_\pm$ as in \eqref{tilCJ}.

\section{The naive \texorpdfstring{$c\to0$}{c to 0} limit of the covariant position operator}
\label{app:bigXlimit}

It is natural to ask whether the observable coordinate \eqref{XC} can
also be reached by letting $c\to0$ directly in the relativistic
covariant (Foldy--Wouthuysen-type) position \eqref{bigX}, rather than
from the boost \eqref{exoboost}.  Expanding \eqref{bigX} in components
with the convention $\epsilon^{012}=1$ gives, for the spatial directions,
\be
X_1=x_1+(P^\rho P_\rho)^{-1}
\bigl(P^{0}\CJ^{2}-P^{2}\CJ^{0}\bigr)\,,\qquad
X_2=x_2+(P^\rho P_\rho)^{-1}
\bigl(-P^{0}\CJ^{1}+P^{1}\CJ^{0}\bigr)\,,
\ee
and therefore, 
\be
\label{bigXpm}
X_\pm=x_\pm\mp i\,(P^\rho P_\rho)^{-1}
\bigl(P^{0}\CJ_\pm+P_\pm\CJ_0\bigr)\,.
\ee
Substituting the same exotic-limit values used to obtain
\eqref{exoboost} from \eqref{Kpm}, namely $P^{0}\to P_t/c$,
$c\CJ_\pm=\sqrt{2\kappa_C}\,\,
\widetilde b_{\kappa_C/c^2}^{\,\pm}
\to\sqrt{2\kappa_C}\,b^{\pm}$,
$\CJ_0=\alpha+\N\to\kappa_C/c^{2}+\N$, and
$(P^\rho P_\rho)^{-1}
=[-(P^{0})^{2}+P_+P_-]^{-1}\to-c^{2}/P_t^{2}$, both terms in
\eqref{bigXpm} survive at the same order and combine, as $c\to0$, into a
\emph{finite} limit for both $\pm$ components alike:
\be
\label{bigXlim}
\lim_{c\to0}X_\pm=\mathscr X_\pm\pm i\kappa_C P_t^{-2}P_\pm\,,
\ee
with $\mathscr X_\pm$ the off-shell lift \eqref{Xoff}.  The limit is finite
and reproduces it up to an extra term linear in the momentum.
Because $[P_i,P_j]=0$, this term commutes with every momentum and so
leaves the canonical relations $[\mathscr X_i,P_j]=i\delta_{ij}$
untouched.  In real components it is the Bopp shift
$X_i^{\mathrm{lim}}:=\lim_{c\to0}X_i
=\mathcal X_i-\theta_C\epsilon_{ij}P_j$ on shell, and therefore it
reverses the spatial commutator,
$[X_1^{\mathrm{lim}},X_2^{\mathrm{lim}}]=+i\theta_C$, relative to
\eqref{XCnc}.  On shell, $P_t=E$, so the additional term in \eqref{bigXlim} becomes
$\pm i\kappa_C P_t^{-2}P_\pm
=\pm i\theta_C P_\pm$, with $\theta_C=\frac{\kappa_C}{E^2}$ the noncommutativity parameter of
\eqref{XCnc}. This is a momentum-dependent, Bopp-shift-type redefinition of the noncommutative coordinate, of the kind familiar from the freedom to reorder or redefine coordinates on a noncommutative plane.  What matters is that the naive limit of
\eqref{bigX} does \emph{not} coincide with $\mathcal X_\pm$ on shell, but with a
Bopp-shifted version of it.  This is why the main text constructs
$\mathcal X_\pm$ directly from the already-contracted boost
\eqref{exoboost} rather than from \eqref{bigX}.

\bibliographystyle{unsrt}
\bibliography{carroll-anyons}

@misc{Levy-Leblond:1965dsc,
  note = {J.~M.~L\'evy-Leblond,
``Une nouvelle limite non-relativiste du groupe de Poincar\'e,''
Ann.\ Inst.\ H.\ Poincar\'e Phys.\ Theor.\ \textbf{3} (1965) 1.}
}

@misc{SenGupta:1966qer,
  note = {N.~D.~Sen Gupta,
``On an analogue of the Galilei group,''
Nuovo Cim.\ A \textbf{44} (1966) 512.}
}

@misc{Bacry:1968zf,
  note = {H.~Bacry and J.~L\'evy-Leblond,
``Possible kinematics,''
J.\ Math.\ Phys.\ \textbf{9} (1968) 1605.}
}

@misc{Duval:2014uoa,
  note = {C.~Duval, G.~W.~Gibbons, P.~A.~Horvathy and P.~M.~Zhang,
``Carroll versus Newton and Galilei: two dual non-Einsteinian concepts of
time,''
Class.\ Quant.\ Grav.\ \textbf{31} (2014) 085016
[arXiv:1402.0657].}
}

@misc{Ciambelli:2019lap,
  note = {L.~Ciambelli, R.~G.~Leigh, C.~Marteau and P.~M.~Petropoulos,
``Carroll structures, null geometry and conformal isometries,''
Phys.\ Rev.\ D \textbf{100} (2019) 046010
[arXiv:1905.02221].}
}

@misc{Donnay:2022aba,
  note = {L.~Donnay, A.~Fiorucci, Y.~Herfray and R.~Ruzziconi,
``Carrollian perspective on celestial holography,''
Phys.\ Rev.\ Lett.\ \textbf{129} (2022) 071602
[arXiv:2202.04702].}
}

@misc{Mason:2023mti,
  note = {L.~Mason, R.~Ruzziconi and A.~Yelleshpur Srikant,
``Carrollian amplitudes and celestial symmetries,''
JHEP \textbf{05} (2024) 012
[arXiv:2312.10138].}
}

@misc{Alday:2024yyj,
  note = {L.~F.~Alday, M.~Nocchi, R.~Ruzziconi and A.~Yelleshpur Srikant,
``Carrollian amplitudes from holographic correlators,''
JHEP \textbf{03} (2025) 158
[arXiv:2406.19343].}
}

@misc{Donnay:2019jiz,
  note = {L.~Donnay and C.~Marteau,
``Carrollian physics at the black hole horizon,''
Class.\ Quant.\ Grav.\ \textbf{36} (2019) 165002
[arXiv:1903.09654].}
}

@misc{Ecker:2023uwm,
  note = {F.~Ecker, D.~Grumiller, J.~Hartong, A.~P\'erez, S.~Prohazka and
R.~Troncoso,
``Carroll black holes,''
SciPost Phys.\ \textbf{15} (2023) 245
[arXiv:2308.10947].}
}

@misc{Marsot:2022imf,
  note = {L.~Marsot, P.~M.~Zhang, M.~Chernodub and P.~A.~Horvathy,
``Hall effects in Carroll dynamics,''
Phys.\ Rept.\ \textbf{1028} (2023) 1
[arXiv:2212.02360].}
}

@misc{deBoer:2021jej,
  note = {J.~de Boer, J.~Hartong, N.~A.~Obers, W.~Sybesma and S.~Vandoren,
``Carroll symmetry, dark energy and inflation,''
Front.\ in Phys.\ \textbf{10} (2022) 810405
[arXiv:2110.02319].}
}

@misc{Isberg:1993av,
  note = {J.~Isberg, U.~Lindstrom, B.~Sundborg and G.~Theodoridis,
``Classical and quantized tensionless strings,''
Nucl.\ Phys.\ B \textbf{411} (1994) 122
[arXiv:hep-th/9307108].}
}

@misc{Bagchi:2015nca,
  note = {A.~Bagchi, S.~Chakrabortty and P.~Parekh,
``Tensionless strings from worldsheet symmetries,''
JHEP \textbf{01} (2016) 158
[arXiv:1507.04361].}
}

@misc{Klauder:1970cs,
  note = {J.~R.~Klauder,
``Ultralocal scalar field models,''
Commun.\ Math.\ Phys.\ \textbf{18} (1970) 307.}
}

@misc{Bergshoeff:2015sic,
  note = {E.~Bergshoeff, J.~Rosseel and T.~Zojer,
``Non-relativistic fields from arbitrary contracting backgrounds,''
Class.\ Quant.\ Grav.\ \textbf{33} (2016) 175010
[arXiv:1512.06064].}
}

@misc{Bagchi:2016bcd,
  note = {A.~Bagchi, R.~Basu, A.~Kakkar and A.~Mehra,
``Flat holography: aspects of the dual field theory,''
JHEP \textbf{12} (2016) 147
[arXiv:1609.06203].}
}

@misc{Bergshoeff:2017btm,
  note = {E.~Bergshoeff, J.~Gomis, B.~Rollier, J.~Rosseel and T.~ter Veldhuis,
``Carroll versus Galilei gravity,''
JHEP \textbf{03} (2017) 165
[arXiv:1701.06156].}
}

@misc{Hansen:2021fxi,
  note = {D.~Hansen, N.~A.~Obers, G.~Oling and B.~T.~S{\o}gaard,
``Carroll expansion of general relativity,''
SciPost Phys.\ \textbf{13} (2022) 055
[arXiv:2112.12684].}
}

@misc{Bergshoeff:2022eog,
  note = {E.~Bergshoeff, J.~Figueroa-O'Farrill and J.~Gomis,
``A non-Lorentzian primer,''
SciPost Phys.\ Lect.\ Notes \textbf{69} (2023) 1
[arXiv:2206.12177].}
}

@misc{Bagchi:2019xfx,
  note = {A.~Bagchi, A.~Mehra and P.~Nandi,
``Field theories with conformal Carrollian symmetry,''
JHEP \textbf{05} (2019) 108
[arXiv:1901.10147].}
}

@misc{Bergshoeff:2023vfd,
  note = {E.~A.~Bergshoeff, A.~Campoleoni, A.~Fontanella, L.~Mele and J.~Rosseel,
``Carroll fermions,''
SciPost Phys.\ \textbf{16} (2024) 153
[arXiv:2312.00745].}
}

@misc{Koutrolikos:2023evq,
  note = {K.~Koutrolikos and M.~Najafizadeh,
``Super-Carrollian and super-Galilean field theories,''
Phys.\ Rev.\ D \textbf{108} (2023) 125014
[arXiv:2309.16786].}
}

@misc{Ekiz:2025hdn,
  note = {E.~Ekiz, E.~O.~Kahya and U.~Zorba,
``Quantization of Carrollian fermions,''
Phys.\ Rev.\ D \textbf{111} (2025) 105019
[arXiv:2502.05645].}
}

@misc{Marsot:2021tvq,
  note = {L.~Marsot,
  ``Planar Carrollean dynamics, and the Carroll quantum equation,''
  J.\ Geom.\ Phys.\ \textbf{179} (2022) 104574
  [arXiv:2110.08489].}
}

@misc{deAzcarraga:1997mdw,
  note = {J.~A.~de Azcarraga, F.~J.~Herranz, J.~C.~Perez Bueno and
  M.~Santander,
  ``Central extensions of the quasiorthogonal Lie algebras,''
  J.\ Phys.\ A \textbf{31} (1998) 1373
  [arXiv:q-alg/9612021].}
}

@misc{Matulich:2019cdo,
  note = {J.~Matulich, S.~Prohazka and J.~Salzer,
  ``Limits of three-dimensional gravity and metric kinematical Lie
  algebras in any dimension,''
  JHEP \textbf{07} (2019) 118
  [arXiv:1903.09165].}
}

@misc{Mehra:2023rmm,
  note = {A.~Mehra and A.~Sharma,
  ``Toward Carrollian quantization: Renormalization of Carrollian
  electrodynamics,''
  Phys.\ Rev.\ D \textbf{108} (2023) 046019
  [arXiv:2302.13257].}
}

@misc{Ciambelli:2023xqk,
  note = {L.~Ciambelli,
  ``Dynamics of Carrollian scalar fields,''
  Class.\ Quant.\ Grav.\ \textbf{41} (2024) 165011
  [arXiv:2311.04113].}
}

@misc{Sharma:2025rug,
  note = {A.~Sharma,
  ``Studies on Carrollian quantum field theories,''
  Class.\ Quant.\ Grav.\ \textbf{43} (2026) 045006
  [arXiv:2502.00487].}
}

@misc{Zeng:2024bcl,
  note = {H.~X.~Zeng, Q.~L.~Zhao, P.~M.~Zhang and P.~A.~Horvathy,
``Peierls substitution and Hall motion in exotic Carroll dynamics,''
Phys.\ Rev.\ D \textbf{111} (2025) 044025
[arXiv:2411.14329].}
}

@misc{Brink:2002zx,
  note = {L.~Brink, A.~M.~Khan, P.~Ramond and X.~Xiong,
``Continuous spin representations of the Poincar\'e and super-Poincar\'e
groups,''
J.\ Math.\ Phys.\ \textbf{43} (2002) 6279
[arXiv:hep-th/0205145].}
}

@misc{Khan:2004nj,
  note = {A.~M.~Khan and P.~Ramond,
``Continuous spin representations from group contraction,''
J.\ Math.\ Phys.\ \textbf{46} (2005) 053515
[Erratum: J.\ Math.\ Phys.\ \textbf{46} (2005) 079901]
[arXiv:hep-th/0410107].}
}

@misc{Bekaert:2017khg,
  note = {X.~Bekaert and E.~D.~Skvortsov,
``Elementary particles with continuous spin,''
Int.\ J.\ Mod.\ Phys.\ A \textbf{32} (2017) 1730019
[arXiv:1708.01030].}
}

@misc{Figueroa-OFarrill:2023qty,
  note = {J.~Figueroa-O'Farrill, A.~P\'erez and S.~Prohazka,
``Quantum Carroll/fracton particles,''
JHEP \textbf{10} (2023) 041
[arXiv:2307.05674].}
}

@misc{Banerjee:2023jpi,
  note = {K.~Banerjee, R.~Basu, B.~Krishnan, S.~Maulik, A.~Mehra and A.~Ray,
``One-loop quantum effects in Carroll scalars,''
Phys.\ Rev.\ D \textbf{108} (2023) 085022
[arXiv:2307.03901].}
}

@misc{Aldaya:1985gy,
  note = {V.~Aldaya and J.~A.~de Azc\'arraga,
``Group manifold analysis of the structure of relativistic quantum
mechanics,''
Annals Phys.\ \textbf{165} (1985) 484.}
}

@misc{Duval:2002cw,
  note = {C.~Duval and P.~A.~Horvathy,
``Spin and exotic Galilean symmetry,''
Phys.\ Lett.\ B \textbf{547} (2002) 306
[Erratum: Phys.\ Lett.\ B \textbf{588} (2004) 228]
[arXiv:hep-th/0209166].}
}

@misc{Leinaas:1977fm,
  note = {J.~M.~Leinaas and J.~Myrheim,
``On the theory of identical particles,''
Nuovo Cim.\ B \textbf{37} (1977) 1.}
}

@misc{Wilczek:1982wy,
  note = {F.~Wilczek,
``Quantum mechanics of fractional spin particles,''
Phys.\ Rev.\ Lett.\ \textbf{49} (1982) 957.}
}

@misc{Wu:1984hj,
  note = {Y.~S.~Wu,
``General theory for quantum statistics in two dimensions,''
Phys.\ Rev.\ Lett.\ \textbf{52} (1984) 2103.}
}

@misc{Tsui:1982yy,
  note = {D.~C.~Tsui, H.~L.~Stormer and A.~C.~Gossard,
``Two-dimensional magnetotransport in the extreme quantum limit,''
Phys.\ Rev.\ Lett.\ \textbf{48} (1982) 1559.}
}

@misc{Laughlin:1983fy,
  note = {R.~B.~Laughlin,
``Anomalous quantum Hall effect: an incompressible quantum fluid with
fractionally charged excitations,''
Phys.\ Rev.\ Lett.\ \textbf{50} (1983) 1395.}
}

@misc{Halperin:1984fn,
  note = {B.~I.~Halperin,
``Statistics of quasiparticles and the hierarchy of fractional quantized
Hall states,''
Phys.\ Rev.\ Lett.\ \textbf{52} (1984) 1583.}
}

@misc{Arovas:1984qr,
  note = {D.~Arovas, J.~R.~Schrieffer and F.~Wilczek,
``Fractional statistics and the quantum Hall effect,''
Phys.\ Rev.\ Lett.\ \textbf{53} (1984) 722.}
}

@misc{dePicciotto:1997qc,
  note = {R.~de Picciotto, M.~Reznikov, M.~Heiblum, V.~Umansky, G.~Bunin and
D.~Mahalu,
``Direct observation of a fractional charge,''
Nature \textbf{389} (1997) 162
[arXiv:cond-mat/9707289].}
}

@misc{Bartolomei:2020qfr,
  note = {H.~Bartolomei \textit{et al.},
``Fractional statistics in anyon collisions,''
Science \textbf{368} (2020) 173
[arXiv:2006.13157].}
}

@misc{Nakamura:2020mok,
  note = {J.~Nakamura, S.~Liang, G.~C.~Gardner and M.~J.~Manfra,
``Direct observation of anyonic braiding statistics,''
Nature Phys.\ \textbf{16} (2020) 931
[arXiv:2006.14115].}
}

@misc{Kitaev:1997wr,
  note = {A.~Y.~Kitaev,
``Fault tolerant quantum computation by anyons,''
Annals Phys.\ \textbf{303} (2003) 2
[arXiv:quant-ph/9707021].}
}

@misc{Kitaev:2005hzj,
  note = {A.~Kitaev,
``Anyons in an exactly solved model and beyond,''
Annals Phys.\ \textbf{321} (2006) 2
[arXiv:cond-mat/0506438].}
}

@misc{Nayak:2008zza,
  note = {C.~Nayak, S.~H.~Simon, A.~Stern, M.~Freedman and S.~Das Sarma,
``Non-Abelian anyons and topological quantum computation,''
Rev.\ Mod.\ Phys.\ \textbf{80} (2008) 1083
[arXiv:0707.1889].}
}

@misc{Andersen:2022xmz,
  note = {T.~I.~Andersen \textit{et al.} [Google Quantum AI],
``Non-Abelian braiding of graph vertices in a superconducting processor,''
Nature \textbf{618} (2023) 264
[arXiv:2210.10255].}
}

@misc{Boulanger:2013naa,
  note = {N.~Boulanger, P.~Sundell and M.~Valenzuela,
``Three-dimensional fractional-spin gravity,''
JHEP \textbf{02} (2014) 052
[arXiv:1312.5700].}
}

@misc{Boulanger:2015uha,
  note = {N.~Boulanger, P.~Sundell and M.~Valenzuela,
``Gravitational and gauge couplings in Chern-Simons fractional spin
gravity,''
JHEP \textbf{01} (2016) 173
[arXiv:1504.04286].}
}

@misc{Jackiw:1990ka,
  note = {R.~Jackiw and V.~P.~Nair,
``Relativistic wave equations for anyons,''
Phys.\ Rev.\ D \textbf{43} (1991) 1933.}
}

@misc{Plyushchay:1990cv,
  note = {M.~S.~Plyushchay,
``Relativistic model of the anyon,''
Phys.\ Lett.\ B \textbf{248} (1990) 107.}
}

@misc{Plyushchay:1991qd,
  note = {M.~S.~Plyushchay,
``Fractional spin: Majorana--Dirac field,''
Phys.\ Lett.\ B \textbf{273} (1991) 250.}
}

@misc{Plyushchay:1990rt,
  note = {M.~S.~Plyushchay,
``Relativistic particle with torsion, Majorana equation and fractional
spin,''
Phys.\ Lett.\ B \textbf{262} (1991) 71.}
}

@misc{Volkov:1989qa,
  note = {D.~V.~Volkov,
``Quartions in relativistic field theory,''
JETP Lett.\ \textbf{49} (1989) 541
[Pisma Zh.\ Eksp.\ Teor.\ Fiz.\ \textbf{49} (1989) 473].}
}

@misc{Sorokin:1992sy,
  note = {D.~P.~Sorokin and D.~V.~Volkov,
``(Anti)commuting spinors and supersymmetric dynamics of semions,''
Nucl.\ Phys.\ B \textbf{409} (1993) 547.}
}

@misc{Cortes:1992fa,
  note = {J.~L.~Cort\'es and M.~S.~Plyushchay,
``Linear differential equations for a fractional spin field,''
J.\ Math.\ Phys.\ \textbf{35} (1994) 6049
[arXiv:hep-th/9405193].}
}

@misc{Cortes:1995wa,
  note = {J.~L.~Cort\'es and M.~S.~Plyushchay,
``Anyons as spinning particles,''
Int.\ J.\ Mod.\ Phys.\ A \textbf{11} (1996) 3331
[arXiv:hep-th/9505117].}
}

@misc{Bargmann:1946me,
  note = {V.~Bargmann,
``Irreducible unitary representations of the Lorentz group,''
Annals Math.\ \textbf{48} (1947) 568.}
}

@misc{Barut:1965,
  note = {A.~O.~Barut and C.~Fronsdal,
``On non-compact groups. II. Representations of the $2+1$ Lorentz group,''
Proc.\ Roy.\ Soc.\ Lond.\ A \textbf{287} (1965) 532.}
}

@misc{Deser:1981wh,
  note = {S.~Deser, R.~Jackiw and S.~Templeton,
``Topologically massive gauge theories,''
Annals Phys.\ \textbf{140} (1982) 372.}
}

@misc{Levy-Leblond:1967eic,
  note = {J.~M.~L\'evy-Leblond,
``Nonrelativistic particles and wave equations,''
Commun.\ Math.\ Phys.\ \textbf{6} (1967) 286.}
}

@misc{Lukierski:1996br,
  note = {J.~Lukierski, P.~C.~Stichel and W.~J.~Zakrzewski,
``Galilean-invariant $(2+1)$-dimensional models with a Chern-Simons-like
term and $D=2$ noncommutative geometry,''
Annals Phys.\ \textbf{260} (1997) 224
[arXiv:hep-th/9612017].}
}

@misc{Jackiw:2000tz,
  note = {R.~Jackiw and V.~P.~Nair,
``Anyon spin and the exotic central extension of the planar Galilei
group,''
Phys.\ Lett.\ B \textbf{480} (2000) 237
[arXiv:hep-th/0003130].}
}

@misc{Duval:2000xr,
  note = {C.~Duval and P.~A.~Horvathy,
``The exotic Galilei group and the `Peierls substitution',''
Phys.\ Lett.\ B \textbf{479} (2000) 284
[arXiv:hep-th/0002233].}
}

@misc{Horvathy:2004fw,
  note = {P.~A.~Horvathy and M.~S.~Plyushchay,
``Anyon wave equations and the noncommutative plane,''
Phys.\ Lett.\ B \textbf{595} (2004) 547
[arXiv:hep-th/0404137].}
}

@misc{Horvathy:2002vt,
  note = {P.~A.~Horvathy and M.~S.~Plyushchay,
``Non-relativistic anyons, exotic Galilean symmetry and noncommutative
plane,''
JHEP \textbf{06} (2002) 033
[arXiv:hep-th/0201228].}
}

@misc{Duval:2001hu,
  note = {C.~Duval and P.~A.~Horvathy,
``Exotic Galilean symmetry in the non-commutative plane, and the Hall
effect,''
J.\ Phys.\ A \textbf{34} (2001) 10097
[arXiv:hep-th/0106089].}
}

@misc{Foldy:1949wa,
  note = {L.~L.~Foldy and S.~A.~Wouthuysen,
``On the Dirac theory of spin 1/2 particles and its non-relativistic
limit,''
Phys.\ Rev.\ \textbf{78} (1950) 29.}
}

@misc{Horvathy:2006pw,
  note = {P.~A.~Horvathy, M.~S.~Plyushchay and M.~Valenzuela,
``Bosonized supersymmetry of anyons and supersymmetric exotic particle on
the non-commutative plane,''
Nucl.\ Phys.\ B \textbf{768} (2007) 247
[arXiv:hep-th/0610317].}
}

@misc{Horvathy:2010vm,
  note = {P.~A.~Horvathy, M.~S.~Plyushchay and M.~Valenzuela,
``Bosons, fermions and anyons in the plane, and supersymmetry,''
Annals Phys.\ \textbf{325} (2010) 1931
[arXiv:1001.0274].}
}

@misc{Parekh:2026wri,
  note = {P.~Parekh, A.~Sharma and M.~Valenzuela,
``Interacting Galilean and finite-energy Carroll fermions,''
[arXiv:2608.05324 [hep-th]].}
}

@misc{Campoleoni:2021blr,
  note = {A.~Campoleoni and S.~Pekar,
``Carrollian and Galilean conformal higher-spin algebras in any dimensions,''
JHEP \textbf{02} (2022) 150
[arXiv:2110.07794].}
}

@misc{Liu:2023jnc,
  note = {W.~B.~Liu, J.~Long and X.~H.~Zhou,
``Quantum flux operators in higher spin theories,''
Phys.\ Rev.\ D \textbf{109} (2024) 086012
[arXiv:2311.11361].}
}

@misc{Bergshoeff:2014jla,
  note = {E.~Bergshoeff, J.~Gomis and G.~Longhi,
``Dynamics of Carroll particles,''
Class.\ Quant.\ Grav.\ \textbf{31} (2014) 205009
[arXiv:1405.2264].}
}

@misc{Bopp:1956,
  note = {F.~Bopp,
``La m\'ecanique quantique est-elle une m\'ecanique statistique classique particuli\`ere?,''
Ann.\ Inst.\ H.\ Poincar\'e \textbf{15} (1956) 81--112.}
}

@misc{Horvathy:2002wc,
  note = {P.~A.~Horvathy,
``The noncommutative Landau problem and the Peierls substitution,''
Annals Phys.\ \textbf{299} (2002) 128
[arXiv:hep-th/0201007].}
}

\end{document}